\documentclass[journal]{IEEEtran}

\usepackage{amsmath}   %
\usepackage[psamsfonts]{amssymb}
\interdisplaylinepenalty=2500\usepackage{type1cm}
\usepackage{nicefrac}
\usepackage{xspace}
\usepackage{cite}
\usepackage{graphicx}
\usepackage{subfigure}
\newcommand{\aposteriori}{\textit{a~posteriori}\xspace} %
\newcommand{\apriori}{\textit{a~priori}\xspace} %
\def\ie{\textrm{i.e.}\xspace} %
\def\Hline{\noalign{\hrule height 0.2mm}} %
\makeatletter
\def\putfloat{\@testopt\@putfloat{}}
\def\@putfloat[#1]#2{%
   \@ifundefined{@FC@#2}%
      {\typeout{<<<Float-id ``#2'' is not defined.>>>}%
       \G@refundefinedtrue}%
      {\gdef\@curr@fps{#1}%
       \@nameuse{@FC@#2}%
       \global\let\@curr@fps\@empty}}
\def\savefloat#1{%
   \edef\@curr@floatid{#1}%
   \let\do\@makeother \dospecials
   \def\par{^^J}%
   \obeylines
   \@savefloat}
\global\let\@curr@fps\@empty
\let\@@xfloat\@xfloat
\def\@xfloat#1[#2]{%
   \ifx\@curr@fps\@empty
      \protected@edef\@tempa{\noexpand#1[#2]}%
   \else
      \protected@edef\@tempa{\noexpand#1[\@curr@fps]}%
   \fi
   \expandafter\@@xfloat\@tempa}
\begingroup
\catcode`\[=1  \catcode`\]=2
\catcode`\{=12 \catcode`\}=12
\catcode`\|=0  \catcode`\\=12
|gdef|@savefloat#1\end{savefloat}[%
   |def|@curr@float@contents[#1]%
   |end[savefloat]]
|endgroup
\def\endsavefloat{%
   \if@filesw
      \immediate\write\@auxout{%
         \string\FloatContents{\@curr@floatid}{\@curr@float@contents}}%
   \fi}
\def\@tempa{\long\def\@newl@bel##1##2##3}
\expandafter\@tempa\expandafter{\@newl@bel{#1}{#2}{#3}}
\def\@tempa{\long\def\@testdef##1##2##3}
\expandafter\@tempa\expandafter{\@testdef{#1}{#2}{#3}}
\def\FloatContents{\@newl@bel{@FC}}
\makeatother%
\begin{document}

\title{The accurate optimal-success/error-rate calculations applied to the realizations of the reliable and short-period integer ambiguity resolution in carrier-phase GPS/GNSS positioning}
\author{Kentaro~Kondo}

\maketitle\begin{abstract}

The maximum-marginal-\textit{a-posteriori} success rate of statistical decision under multivariate Gaussian error distribution on an integer lattice is almost rigorously calculated by using union-bound approximation and Monte Carlo integration. These calculations are applied to the revelation of the various possible realizations of the reliable and short-period integer ambiguity resolution in precise carrier-phase relative positioning by GPS/GNSS. The theoretical foundation and efficient methodology are systematically developed, and two types of the enhancement of union-bound approximation are proposed and examined.

The results revealed include an extremely high reliability under the condition of accurate carrier-phase measurements and a large number of visible satellites, its heavy degradation caused by the slight amount of differentiated ionospheric delays due to the nonvanishing baseline length between rover and reference receivers, and the advantages of the use of the multiple carrier frequencies. The succeeding initialization of the integer ambiguities is shown to overcome the disadvantageous condition of the nonvanishing baseline length effectively due to the reasonably assumed temporal and spatial constancy of differentiated ionospheric delays. 
\end{abstract}

\begin{keywords}
Bayes decision, lattice basis reduction, Fincke-Pohst algorithm, union bound, Monte Carlo integration, GPS, GNSS, carrier-phase positioning, integer ambiguity resolution, ionospheric delay.
\end{keywords}
\IEEEpeerreviewmaketitle%
%

\PARstart{C}{orrect} resolution of integer ambiguities is indispensable for precise carrier-phase relative positioning with the Global Positioning System (GPS) and the global navigation satellite system (GNSS) \cite{LeickA2004, HofmannWellenhoffBetal2001,TeunissenPJGetKleusbergA1998}. This means that one needs an extremely high success rate and an extremely low error rate of the resolution, even in the heavily difficult resolution for non-permanent, or mobile, receivers used in realistic and unstable environments \cite{KondoK2001}.

The relative positioning requires rover and reference receivers, and precisely measures the differentiated coordinates of the position, \ie{}, the baseline vector spanned between the positions of the two receivers. Consequently, the environmental conditions for the rover receiver are crucial for the ambiguity resolution in normal cases.

In practice, the improvement in the ambiguity resolution is inductively known to be achieved by using accurate carrier-phase and code-pseudorange measurements, a large number of visible satellites, multiple carrier frequencies, and other means.
Although such conditions improve the statistical success and error rates, their reliable estimation seems not to have been thoroughly investigated. This is partially due to the fact that the theoretical calculations are extremely laborious high-dimensional integrations of a multivariate error probability distribution function in a complicated integration domain.

The theoretical problems of the ambiguity resolution are formalized as optimal statistical decisions, and these are classified as either fixed-sample decisions or sequential ones according to the treatment of time series of the measurements, \ie{}, statistical samples. The first kind of decision uses assigned fixed-number epochs of measurements and generally chooses between making and avoiding a judgment. The second kind of decision is allowed to re-sample the measurement data sequentially. If a judgment is made, its optimal decision should choose the maximum marginal \aposteriori{} solution for both kinds of decisions according to the Bayes statistical theory \cite{BergerJO1985}. It is identical to the least-square solution in the case of the simple linear and Gaussian error distribution model.Several studies in the field of GPS/GNSS have discussed the theoretically optimal fixed-sample resolutions of integer ambiguities and the calculations of the success rate \cite{TeunissenPJG1999c,HassibiAetBoydS1998,ThomsenHF2000}. Other studies have discussed the sequential resolutions and the calculations of the success rate \cite{MacabiauCetBenhallamA1996,WolfeJDetal2001,PolingTCetZatezaloA2002}. Additional studies have described the efficient computational algorithms to search the integer least-square solution \cite{TeunissenPJG1995, HassibiAetBoydS1998,AbdelHafezMFetal2002,GrafarendEWP2000}. However, neither the rigorous aspect of statistical optimality nor the depth calculation of success and error rates seems to have been thoroughly investigated as yet \cite{KondoK2003a} in the field of GPS/GNSS.

Although the sequential resolution is required for the real-time positioning of mobile receivers, this study concentrates on the fixed-sample resolution, because it is required for the off-line data processing, and, furthermore, is considered to be fundamental to the study of the sequential one.

Various conditions which are intended to improve the performance of the integer ambiguity resolution have been discussed in several studies. Such conditions include the use of precise code-pseudorange and carrier-phase measurements \cite{LachapelleGetal1992, HanSetRizosC1996}, the short baseline length of relative positioning \cite{PrattMetal1997}, the use of a reference receiver network \cite{WanningerL1995,RaquetJ1998,ColomboOLetal1999}, the use of multiple carrier-phase measurements, or carrier frequencies \cite{HatchR1996}, more visible satellites, including planned European Galileo navigation ones \cite{HeinGWetal2001,EissfellerBetal2002}, the preferable satellites' constellation and movement \cite{TeunissenPJGetOdijkD1997}, and longer sampling duration \cite{HanSetJohnsonR2001}. Most of these investigations, however, do not seem to touch upon the rigorous argument on the performance of the integer ambiguity resolution, which should be based on exact statistical and mathematical discussions.This study systematically discusses the theoretical foundation and efficient methodology of the integer ambiguity resolution as formalized to be an optimal fixed-sample statistical decision, and develops the almost rigorous calculation of the maximum-marginal-\textit{a-posteriori} success/error rates under multivariate Gaussian error distribution on an integer lattice by using union-bound \cite{WozencraftJMetJacobsIM1965,ConwayJHetSloaneNJA1999,ProakisJG2000} approximation and importance-sampling Monte Carlo integration \cite{HammersleyJMetHandscombDC1964}.

This study also proposes and examines the two types of the enhancement of union-bound approximation of the rates for such optimal decisions that are allowed to avoid a judgment and the optimal decision that addresses a marginalized integer lattice. The latter type of decision is applied to the succeeding initialization of the integer ambiguity as discussed in Section~II-D.

Highly efficient methods are also scrupulously developed and applied to the exact calculation of the maximum-marginal-\textit{a-posteriori} and (conditional) Bayes decisions, which are required in the process of the Monte Carlo calculations of the rates, by using an improved Fincke-Pohst algorithm \cite{PohstM1981,FinckeUetPohstM1985,CohenH1993}, or sphere decoding.

This study, then, applies these efficient methodologies to calculating the accurate success rates of integer ambiguity resolution and reveals the fundamental aspects of the reliability performance of the short-period integer ambiguity resolution in carrier-phase relative positioning.

The aspects revealed include the extremely high reliability realized by the optimal resolution under the condition of a large number of visible satellites, its heavy degradation caused by the slight amount of differentiated ionospheric delays due to the nonvanishing baseline length of relative positioning, the advantages of the use of the multiple carrier frequencies (GPS-L1, L2, and L5 signals are considered in this study), and their mutual relations associated with the dependence on the error in the carrier-phase and code-pseudorange measurements.

The performance of the succeeding initialization is shown to be greatly improved by the reasonably assumed temporal and spatial constancy of differentiated ionospheric delays, and, furthermore, by the temporal constancy of the receiver's differential clock offset (interfrequency bias) when the continuous measurement of the attitude of the receiver's antenna is available.

The benefit and trade-off of the well-known wide-laning method \cite{CounselmanCCetal1979} using the multiple frequencies are revealed under the condition of the large differentiated ionospheric delays.

The only slight difference is clarified in the performance between the static and kinematic (``on-the-fly'') ambiguity resolutions.\section{GPS/GNSS measurements and integer ambiguities} %

This section describes the statistical formalization of GPS/GNSS measurements and integer ambiguities for the optimal resolution.

\subsection{Basic assumptions} %

As mentioned, we will concentrate on the fixed-sample optimal decision for the integer ambiguity resolution.

The integer ambiguities involved in the carrier-phase measurements are assumed to be initially non-informative, and not to vary during the resolution for the sake of simplicity. No failure in both carrier-phase and code-pseudorange measurements is assumed to occur during the resolution.

All measurements and error variables, including the integer ambiguities, ionospheric delays, the (rover) receiver's coordinates, and the (rover) receiver's clock offset, are assumed to have been single-differentiated, \ie{}, these are the variables relative to the reference receiver's ones. We presume to neglect the actual slight differences in the measuring and transmitting time of the GPS/GNSS signals with regard to the distinction between the rover and reference receivers \cite{LeickA2004}, which is justified for the theoretical discussions, as this study is. The short or medium-range baseline length between the rover and reference receivers is assumed, and is specifically within 30\,km under the conditions of the calculations in Section~V-A.

The GPS/GNSS positioning equations are assumed to be linear, which is justified by the fact that the signal waves are regarded as plane waves when they are received on the earth. GPS/GNSS satellites are assumed to be equivalent in regard to both their nominal signal frequencies and the receiver's clock offsets. The former equivalence is justified in the case of GPS \cite{SpilkerJrJJ1996,ISGPS200D2004} and Galileo \cite{HeinGWetal2001}, and the latter is justified for most modern GPS/GNSS receivers.

Each measurement or prediction error is assumed to be distributed according to a Gaussian function with a vanishing mean and a certain covariance. Gaussian error distributions strongly assume that huge errors or large noises in the measurements will be relatively rare. Furthermore, a stationary random process for the measurement error is assumed, and it is modeled as a combination of a temporally constant component and a temporally varying autoregressive (AR) process component, the latter of which is assumed to be stationary and first-order, \ie{}, AR(1).

The \apriori{} distribution of the (single-differentiated) ionospheric delays is also presumed. It is furthermore presumed to have a vanishing mean and a certain error variance, the latter of which is also presumed to be proportional to the baseline length between rover and reference receivers. The (single-differentiated) ionospheric delays are assumed to be temporally constant during the resolution, and their spatial constancy is also assumed when investigating the improved performance of succeeding initialization. Furthermore, the (single-differentiated) tropospheric delays are assumed to vanish for the sake of simplicity. The (single-differentiated) higher-order terms \cite{BassiriSetHajjGA1993,KedarSetal2003} in ionospheric delay are also assumed to vanish. These assumptions are satisfactorily justified under the condition of the short or medium-range baseline length.

The satellites' constellation and its movement is calculated by using GPS broadcast ephemerides, which are assumed to be accurate enough to neglect their coordinate errors for the short- or medium-range-baseline relative positioning. The sufficiently accurate line-of-sight directions from the satellites to the receiver are assumed to have been calculated by using the coarse estimation of the position of the receiver, which can be calculated by using the ephemerides and the code-pseudorange measurements on the receiver.

The boresights of both the rover and reference receiver's antennas are assumed to be directed toward the same direction, which is likely the local zenith. The attitude of the rover receiver's antenna is assumed to be accurately and continuously measured by using such devices as gyroscopic sensors when investigating the improved performance of succeeding initialization. The GPS/GNSS electromagnetic signal waves \cite{ISGPS200D2004, HeinGWetal2001} are assumed to be exactly right-hand-circularly polarized \cite{BornMetWolfE1999}.

The variation in the phase center of the satellites' and receivers' antennas, which is a slight dependence on the azimuthal angle of the propagating direction of the GPS/GNSS signals and the distinction among their carrier frequencies, is assumed to be neglected.

The signals of satellite navigation are assumed to be GPS-L1, L2, or L5, the carrier frequencies of which are 1.57542, 1.2276, or 1.17645\,GHz \cite{ISGPS200D2004}, respectively. Each carrier frequency is assumed to correspond to the distinctive signal in a one-to-one relationship for the sake of simplicity. Namely, the multiplexing of signal modulations, such as that of inphase and quadrature component modulations or that of phase-shift-keying \cite{ProakisJG2000} and (planned) binary-offset-carrier \cite{BetzJW2001} modulations, is not considered.
We assume the receiver's differential clock offsets among the distinct kinds of measurements (\ie{}, interfrequency bias) do not vary during the resolution. 

Finally, we assume the initial non-informativeness of the (rover) receiver's coordinates, and the initial non-informativeness and non-time-correlativeness for the values and evolution of the (rover) receiver's clock offsets. Both the distinct conditions of the time-constancy and non-time-correlativeness of the (rover) receiver's coordinates are used for investigating the static and kinematic ambiguity resolution, respectively. It is assumed that we are indifferent to the determination of the receiver's clock offset in the ambiguity resolution.
\subsection{Equations of GPS/GNSS measurements} %

Let $\rho_{q,i}^k$ and $\phi_{q,i}^k$ denote (single-differentiated) GPS/GNSS code-pseudorange and carrier-phase measurements, respectively, where $q$, $i$, and $k$ indicate the kind of signals, the epoch of measurements, and the distinction of the satellites, respectively. They satisfy the following equations \cite{LeickA2004, HofmannWellenhoffBetal2001,TeunissenPJGetKleusbergA1998}:
\begin{align}
\rho_{q,i}^k =& \; \boldsymbol{e}_i^k \cdot \boldsymbol{r}_i + \left( \frac{\lambda_q}{\lambda_\textrm{L1}} \right)^2 I_i^k + c \, \tau_{\rho,q,i} + \delta \rho_{q,i}^k
,
\label{basic_equation_1}
\\
\phi_{q,i}^k =& \; N_q^k + \psi_i
\nonumber
\\
\phantom{\phi_{q,i}^k = \;} 
& + \frac{1}{\lambda_q} \boldsymbol{e}_i^k \cdot \boldsymbol{r}_i - \frac{1}{\lambda_q} \left( \frac{\lambda_q}{\lambda_\textrm{L1}} \right)^2 I_i^k + \frac{c}{\lambda_q} \tau_{\phi,q,i} + \delta \phi_{q,i}^k
,
\label{basic_equation_2}
\end{align}
where $\boldsymbol{r}_i$ denotes the coordinates of a receiver, $I_i^k$ the ionospheric delay, $\tau_{\rho, q, i}$ and $\tau_{\phi, q, i}$ the receiver's code-pseudorange and carrier-phase clock offsets (with dependence on the kind of measurements), $\delta \rho_{q,i}^k$ and $\delta \phi_{q,i}^k$ the code-pseudorange and carrier-phase measurement errors, $N_q^k$ the integer ambiguity, and $\psi_i$ the wind-up phase, and, as already mentioned, they are all single-differentiated. $\phi_{q,i}^k$ and $\psi_i$ are expressed in a unit of cycles. $\lambda_q \ (= c / f_q)$ denotes the wavelength of the carrier wave (and $f_q$ is the frequency). $\boldsymbol{e}_i^k$ denotes the unit line-of-sight vector from the satellite $k$ to the receiver, which, as already mentioned, is assumed to have been determined with sufficient accuracy. $I_i^k$ is measured as the delay length at the frequency of GPS-L1.

The (single-differentiated) wind-up phase \cite{WuJTetal1993}, $\psi_i$, does not depend on the distinction among the satellites based on our assumption of the equivalence of the rover and reference receiver's antennas in regard to their boresight's directions. $\psi_i$ is, thus, equal to the (single-differentiated) right-handed rotation angle of the antenna's attitude around its boresight direction based on our assumption of the right-hand-circularly polarized electromagnetic signal waves of the GPS/GNSS signals.

The kinds of signals, $q$'s, are GPS-L1, L2, or L5, as already mentioned. Two artificially synthesized carrier-phase measurements, wide lane ($\text{L}_\text{W}$) and extra-wide one ($\text{L}_\text{EW}$) \cite{CounselmanCCetal1979,HatchR1996}, are additionally introduced by using the following integer unimodular transformation of the three raw carrier-phase measurements:
\begin{equation}
\begin{pmatrix}
\phi_{\textrm{L1},i}^k \\
\phi_{\textrm{W},i}^k \\
\phi_{\textrm{EW},i}^k 
\end{pmatrix}
=
\begin{pmatrix}
1 & 0 & 0 \\
1 & -1 & 0 \\
0 & 1 & -1
\end{pmatrix}
\begin{pmatrix}
\phi_{\textrm{L1},i}^k \\
\phi_{\textrm{L2},i}^k \\
\phi_{\textrm{L5},i}^k 
\end{pmatrix}
.
\label{wide_lanes}
\end{equation}
These carrier-phase measurements should also be accompanied by the respective integer ambiguities which are transformed by the same matrix as above. $q$ will indicate the distinction of these carrier-phase measurements, or that of their carrier frequencies, when considering the integer ambiguities. 

For each kind of signals, $q$, the set of the integer ambiguities for all visible satellites, which have been symbolized as $k$'s, is linearly decomposed as follows:
\begin{equation}
N_q^k = \Delta N_q^k + N_q^0
,
\label{integer_ambiguity_decomposed}
\end{equation}
where $N_q^0$ denotes a common integer increment and $\Delta N_q^k$'s the integer ambiguities differentiated among $k$'s. Any vector of $\Delta N_q^k$'s is assumed to be perpendicular to that of $N_q^0$ for the sake for simplicity.
$\Delta N_q^k$'s are, thus, doubly differentiated among the satellites and between the rover and reference receivers. Note that the nonuniqueness in the actual representation of $\Delta N_q^k$'s should not affect the \textit{optimal} ambiguity resolution.

$\Delta N_q^k$'s are decision parameters in the ambiguity resolution, whereas $N_q^0$ is a nuisance one, which only causes the common increment to $\tau_{\phi, q, i}$'s among $i$'s and to $I_i^k$'s among $i$'s and $k$'s (\ie{}, their common and temporally constant component for each $q$), and never affects the determination of the positioning coordinates, $\boldsymbol{r}_i$, based on our assumption of the equivalence of satellites in regard to both their nominal signal frequencies and the receiver's clock offsets. 

For each epoch, $i$, the set of receiver's clock offsets for both the code-pseudorange and carrier-phase measurements of all kinds of signals, which have been symbolized as $q$'s, is also linearly decomposed as follows:
\begin{equation}
\begin{aligned}
\tau_{\rho,q,i} &= \Delta \tau_{\rho,q,i} + \tau_i^0
,
\\
\tau_{\phi,q,i} &= \Delta \tau_{\phi,q,i} + \tau_i^0
,
\end{aligned}
\label{clock_offsets_decomposed}
\end{equation}
where $\tau_i^0$ denotes a common increment among all kinds of measurements and $\Delta \tau_{\rho, q, i}$'s and $\Delta \tau_{\phi, q, i}$'s the clock offsets differentiated among the distinct kinds of measurements. Any combined vector of $\Delta \tau_{\rho, q, i}$'s and $\Delta \tau_{\phi, q, i}$'s is also assumed to be perpendicular to the vector of $\tau_i^0$.
$\Delta \tau_{\rho, q, i}$'s and $\Delta \tau_{\phi, q, i}$'s are also known as receiver's differential clock offsets or interfrequency biases \cite{LeickA2004}. Their actual representations are not unique, either.

Furthermore, based on this study's model, each of both kinds of the $i$-th-epoch measurement errors, $\delta \rho_{q,i}^k$ and $\delta \phi_{q,i}^k$, is decomposed as follows:
\begin{align}
\delta \rho_{q,i}^k &= \Delta \delta \rho_{q,i}^k +\delta \rho_{q,0}^k ,
\label{measurement_errors_decomposed_1}
\\
\delta \phi_{q,i}^k &= \Delta \delta \phi_{q,i}^k +\delta \phi_{q,0}^k ,
\label{measurement_errors_decomposed_2}
\end{align}
where each of $\delta \rho_{q,0}^k$ and $\delta \phi_{q,0}^k$ is temporally constant component, and each of $\Delta \delta \rho_{q,i}^k$ and $\Delta \delta \phi_{q,i}^k$ is temporally varying one modeled as an AR(1) process as already mentioned.
Let $\boldsymbol{y}_i$ denote the set of all of the $i$-th-epoch measurements, $\rho_{q,i}^k$'s and $\phi_{q,i}^k$'s, being considered. Then, let $\boldsymbol{y}_{(1,2,\ldots,n)}$ denote the time series of the $n$-epoch $\boldsymbol{y}_i$'s:
\begin{equation}
\boldsymbol{y}_{(1,2,\ldots,n)} = \left( \boldsymbol{y}_1,\,\boldsymbol{y}_2,\,\ldots,\,\boldsymbol{y}_n \right) .
\nonumber
\end{equation}
The number of epochs, $n$, is assumed to be fixed in the resolution.

The sets or the time series of the unknown parameters or measurement errors that have already appeared will be expressed by using the same notation as above. The integer ambiguities \eqref{integer_ambiguity_decomposed} are then expressed as follows:
\begin{equation}
\boldsymbol{N} = \Delta \boldsymbol{N} + \boldsymbol{N}^0
,
\label{integer_ambiguity_vector_decomposed}
\end{equation}
and the time series of the receiver's clock offsets \eqref{clock_offsets_decomposed} and measurement errors \eqref{measurement_errors_decomposed_1} and \eqref{measurement_errors_decomposed_2} are expressed respectively as follows:
\begin{gather}
\boldsymbol{\tau}_{(1,2,\ldots,n)} = \Delta \boldsymbol{\tau}_{(1,2,\ldots,n)} + (1,\,1,\,\ldots,\,1)^T \negthinspace \times \tau_{(1,2,\ldots,n)}^0
\nonumber
, \\
\delta \boldsymbol{y}_{(1,2,\ldots,n)} = \Delta \delta \boldsymbol{y}_{(1,2,\ldots,n)} + \left( \delta \boldsymbol{y}_0,\,\delta \boldsymbol{y}_0,\,\ldots,\,\delta \boldsymbol{y}_0 \right)
\nonumber
.
\end{gather}

The statistical likelihood of the set of the unknown parameters subject to $\boldsymbol{y}_{(1,2,\ldots,n)}$ equals the conditional distribution of $\boldsymbol{y}_{(1,2,\ldots,n)}$, which is calculated by using \eqref{basic_equation_1} and \eqref{basic_equation_2} as follows:
\begin{equation}
\begin{aligned}
& \, p(\: \boldsymbol{y}_{(1,2,\ldots,n)} \,|\, \Delta \boldsymbol{N},\, \boldsymbol{N}^0,\, \boldsymbol{r}_{(1,2,\ldots,n)},\, \boldsymbol{I}_{(1,2,\ldots,n)}, \\
& \, \phantom{p(\: \boldsymbol{y}_{(1,2,\ldots,n)} \,|\, \Delta \boldsymbol{N},\,} 
\boldsymbol{\tau}_{(1,2,\ldots,n)},\,
\psi_{(1,2,\ldots,n)} ) \\
= & \, p( \delta \boldsymbol{y}_{(1,2,\ldots,n)} )\\
= & \, \int p( \Delta \delta \boldsymbol{y}_{(1,2,\ldots,n)} ) \, p( \delta \boldsymbol{y}_0) \, d \delta \boldsymbol{y}_0 .
\end{aligned}
\nonumber
\end{equation}

The probability distribution of the stationary AR(1) process, $\Delta \delta \boldsymbol{y}_{(1,2,\ldots,n)}$, is calculated by using the following equations \cite{BrockwellPJetDavisRA1991,DurbinJetKoopmanSJ2001}:
\begin{equation}
\begin{aligned}
& p( \Delta \delta \boldsymbol{y}_{(1,2,\ldots,n)} ) \\
& = \left\{ \prod_{i=2}^n p( \Delta \delta \boldsymbol{y}_i \, |\, \Delta \delta \boldsymbol{y}_{(1,2,\ldots,i-1)} ) \right\} \, p( \Delta \delta \boldsymbol{y}_1)
,
\\
& \Delta \delta \boldsymbol{y}_i = A_1 \, \Delta \delta \boldsymbol{y}_{i-1} + \boldsymbol{\epsilon}_i
,
\end{aligned}
\nonumber
\end{equation}
where $A_1$ is an AR(1) coefficient matrix and $\boldsymbol{\epsilon}_i$ is a stationary error. Note that $p( \Delta \delta \boldsymbol{y}_1)$ (= $p( \Delta \delta \boldsymbol{y}_i)$) is calculated by using $\left\langle \negthinspace \left( \Delta \delta \boldsymbol{y}_i \right) \negthinspace \left( \Delta \delta \boldsymbol{y}_i \right)^T \right\rangle = \sum_{j=0}^\infty \left( A_1 \right)^j \negthinspace \left\langle \boldsymbol{\epsilon}_i \, {\boldsymbol{\epsilon}_i}^T \right\rangle \negthinspace \left( {A_1}^T \right)^j$, and $\left\langle \Delta \delta \boldsymbol{y}_i \right\rangle = 0$.

A marginal joint distribution of $\Delta \boldsymbol{N}$ and $\boldsymbol{y}_{(1,2,\ldots,n)}$ is, thus, calculated by integrating over real nuisance parameters \cite{BergerJO1985}, as follows:
\begin{equation}
\begin{aligned}
& p( \Delta \boldsymbol{N},\, \boldsymbol{y}_{(1,2,\ldots,n)} ) \\
= & \iiiint \sum_{\boldsymbol{N}^0} \, p(\, \boldsymbol{y}_{(1,2,\ldots,n)} \,|\, \Delta \boldsymbol{N},\, \boldsymbol{N}^0,\, \boldsymbol{r}_{(1,2,\ldots,n)},\, \\
& \phantom{\smash{\iiiint \sum_{\boldsymbol{N}^0} \, p( }}
 \boldsymbol{I}_{(1,2,\ldots,n)},\, \boldsymbol{\tau}_{(1,2,\ldots,n)},\, \psi_{(1,2,\ldots,n)} )
\\
& \times p( \Delta \boldsymbol{N},\, \boldsymbol{N}^0,\, \boldsymbol{r}_{(1,2,\ldots,n)},\, \boldsymbol{I}_{(1,2,\ldots,n)},\\
& \phantom{\times p( \Delta \boldsymbol{N},\, \boldsymbol{N}^0,\, \boldsymbol{r}_{(1,2,\ldots,n)},\, }
\boldsymbol{\tau}_{(1,2,\ldots,n)},\, \psi_{(1,2,\ldots,n)})
\\
& \times d \boldsymbol{r}_{(1,2,\ldots,n)}\, d \boldsymbol{I}_{(1,2,\ldots,n)}\, d \boldsymbol{\tau}_{(1,2,\ldots,n)}\, d \psi_{(1,2,\ldots,n)}
,
\end{aligned}
\nonumber
\end{equation}
where an \apriori{} distribution,
\begin{equation}
\begin{aligned}
& p( \Delta \boldsymbol{N},\, \boldsymbol{N}^0,\, \boldsymbol{r}_{(1,2,\ldots,n)},\, \boldsymbol{I}_{(1,2,\ldots,n)},\\
& \phantom{ p( \Delta \boldsymbol{N},\, \boldsymbol{N}^0,\, \boldsymbol{r}_{(1,2,\ldots,n)},\, }
\boldsymbol{\tau}_{(1,2,\ldots,n)},\, \psi_{(1,2,\ldots,n)}) , \\
\end{aligned}
\nonumber
\end{equation}
is used.

The marginal joint distribution is factorized without loss of generality as follows:
\begin{equation}
\begin{aligned}
& p( \Delta \boldsymbol{N},\, \boldsymbol{y}_{(1,2,\ldots,n)} )
\\
= \: & p( \Delta \boldsymbol{N}\,|\, \boldsymbol{\nu} ) \times p(\, \boldsymbol{y}_{(1,2,\ldots,n)} )
,
\end{aligned} 
\nonumber
\end{equation}
and the marginal \aposteriori{} distribution of $\Delta \boldsymbol{N}$ is expressed as:
\begin{equation}
p( \Delta \boldsymbol{N}\,|\, \boldsymbol{y}_{(1,2,\ldots,n)} ) = p( \Delta \boldsymbol{N}\,|\, \boldsymbol{\nu} )
,
\nonumber
\end{equation}
where a minimal sufficient statistic \cite{LehmannEL1986,SchervishMJ1995}, $\boldsymbol{\nu}$, is introduced as a function of $\boldsymbol{y}_{(1,2,\ldots,n)}$:
\begin{equation}
\boldsymbol{\nu} = \boldsymbol{\nu}(\, \boldsymbol{y}_{(1,2,\ldots,n)} )
.
\nonumber
\end{equation}
$\boldsymbol{\nu}$ is called a floating solution of the integer ambiguity in the context of the GPS/GNSS carrier-phase positioning.

The marginal \aposteriori{} distribution is, thus, calculated as follows:
\begin{equation}
p( \Delta \boldsymbol{N}\,|\, \boldsymbol{\nu} ) = \frac{p( \Delta \boldsymbol{N}, \, \boldsymbol{\nu} )}{ \sum_{\Delta \boldsymbol{N}} \, p( \Delta \boldsymbol{N}, \, \boldsymbol{\nu} )}
,
\label{marginal_aposteriori_distribution_1}
\end{equation}
where
\begin{equation}
p( \Delta \boldsymbol{N}, \, \boldsymbol{\nu} ) = \sum_{\boldsymbol{N}^0} \, p(\,\boldsymbol{\nu} \,|\, \Delta \boldsymbol{N}, \boldsymbol{N}^0) \, p(\Delta \boldsymbol{N}, \boldsymbol{N}^0)
,
\label{marginal_joint_distribution}
\end{equation}
and, based on our assumption, the \apriori{} distribution, $p(\Delta \boldsymbol{N}, \boldsymbol{N}^0)$ (= $p(\boldsymbol{N})$), is non-informative, or uniform, with regard to both $\Delta \boldsymbol{N}$ and $\boldsymbol{N}^0$. 
The summation in the right-hand side of \eqref{marginal_joint_distribution} implies that the optimal ambiguity resolution should be generally executed on the integer lattice marginalized over $\boldsymbol{N}^0$ as further discussed in the following sections.

The general form of the conditional distribution of $\boldsymbol{\nu}$ ($\in \boldsymbol{R}^p$), or the likelihood of $\boldsymbol{N}$ ($\in \boldsymbol{Z}^p$), is described as:
\begin{equation}
\begin{aligned} 
& \, p(\,\boldsymbol{\nu} \,|\, \boldsymbol{N})
\\
= & \sqrt{\frac{| M |}{\left( 2 \pi \right)^p}} \exp \negmedspace \left\{ - \frac{1}{2} \left( \boldsymbol{\nu} - \boldsymbol{N} \right)^T \negthinspace M \left( \boldsymbol{\nu} - \boldsymbol{N} \right) \right\}
,
\end{aligned} 
\label{general_conditional_distribution}
\end{equation}
where we used our assumptions of linearity and Gaussian error distribution and introduced the inverse covariance matrix, $M$, of $\boldsymbol{\nu}$. $M$ is calculated by the marginalization over the real nuisance parameters, or practically by using several formulas in linear algebra, such as the completion of squares \cite{BrockwellPJetDavisRA1991,DurbinJetKoopmanSJ2001}.

$p$ is the size of $\boldsymbol{\nu}$ and $\boldsymbol{N}$, and the following relations normally hold:
\begin{gather} 
p = p^0 \times \text{number of visible satellites} ,
\nonumber\\
p^0 = \text{number of carrier frequencies considered} .
\nonumber
\end{gather}
The number of visible satellites is seven and that of carrier frequencies is one, two, or three under the conditions of the calculations in Section~V.

Let $\Delta \boldsymbol{\nu}$ and $\boldsymbol{\nu^0}$ denote the same decomposition of $\boldsymbol{\nu}$ as that of $\boldsymbol{N}$ into $\Delta \boldsymbol{N}$ and $\boldsymbol{N}^0$ \eqref{integer_ambiguity_vector_decomposed}:
\begin{equation}
\boldsymbol{\nu} = \Delta \boldsymbol{\nu} + \boldsymbol{\nu}^0
.
\nonumber
\end{equation}
Note that $\Delta \boldsymbol{\nu}$ is, thus, doubly differentiated as well as $\Delta \boldsymbol{N}$.

If there exists the independence between $\Delta \boldsymbol{N}$ and $\boldsymbol{N}^0$ in the marginal joint distribution, the distribution is factorized as follows:
\begin{equation}
p( \Delta \boldsymbol{N}, \, \boldsymbol{N}^0, \, \Delta \boldsymbol{\nu}, \, \boldsymbol{\nu}^0 ) = p( \Delta \boldsymbol{N}, \, \Delta \boldsymbol{\nu}) \, p( \boldsymbol{N}^0,\, \boldsymbol{\nu}^0 \,|\, \Delta \boldsymbol{\nu} ) 
.
\nonumber
\end{equation}
Let us call this case $\Delta \boldsymbol{N}$-\textit{separable} in distinction from the $\Delta \boldsymbol{N}$-\textit{nonseparable} one, the former of which usually arises under the condition of \apriori{} non-informative $\Delta \tau_{ \phi, q, i}$.

In the $\Delta \boldsymbol{N}$-separable case, the minimal sufficient statistic, thus, should not be $\boldsymbol{\nu}$, but $\Delta \boldsymbol{\nu}$, and the marginal \aposteriori{} distribution of $\Delta \boldsymbol{N}$ is calculated as follows:
\begin{align}
& \; p( \Delta \boldsymbol{N} \,|\, \boldsymbol{\nu} )
\nonumber
\\
&= p( \Delta \boldsymbol{N} \,|\, \Delta \boldsymbol{\nu} )
= \frac{p( \Delta \boldsymbol{N}, \, \Delta \boldsymbol{\nu} )}{ \sum_{\Delta \boldsymbol{N}} \, p( \Delta \boldsymbol{N}, \, \Delta \boldsymbol{\nu} )}
,
\label{marginal_aposteriori_distribution_2}
\\
& \; p( \Delta \boldsymbol{N}, \, \Delta \boldsymbol{\nu} ) = p( \Delta \boldsymbol{\nu} \,|\, \Delta \boldsymbol{N}) \, p(\Delta \boldsymbol{N})
,
\label{Delta_conditional_distribution_0}
\\
& \; p( \Delta \boldsymbol{\nu} \,|\, \Delta \boldsymbol{N})
\nonumber
\\
& = \sqrt{\frac{| M_\Delta | \ }{\left( 2 \pi \right)^{p_\Delta}}} \exp \negmedspace \left\{ - \frac{1}{2} \left( \Delta \boldsymbol{\nu} - \Delta \boldsymbol{N} \right)^T \negthinspace M_\Delta \left( \Delta \boldsymbol{\nu} - \Delta \boldsymbol{N} \right) \right\}
,
\label{Delta_conditional_distribution}
\end{align}
where we introduced the inverse covariance matrix, $M_\Delta$, of $\Delta \boldsymbol{\nu}$, and $p_\Delta = p - p^0$, which is the size of $\Delta \boldsymbol{\nu}$ and $\Delta \boldsymbol{N}$.

An intermediate case can also occur under the condition of \apriori{} non-informative $\psi_i$, informative $\Delta \tau_{ \phi, q, i}$, and $p^0 \geq 2$, where the size of the minimal sufficient statistic is $p - 1$.

Although these kinds of conditions affect the size of the minimal sufficient statistic (\ie{}, the rank of $M$) and the actual calculation in the integer ambiguity resolution, we will use the general form \eqref{general_conditional_distribution} in the following discussion.

In the calculations, each uniform or translationally invariant (location-invariant) infinite-interval (\apriori{}) distribution should be carefully defined to be the limit of a finite-interval distribution (see Appendix~I).
\subsection{Succeeding initialization} %

If the correct differentiated integer ambiguity, $\Delta \boldsymbol{N}$, is once resolved on a multiple-frequency rover receiver, the highly accurate estimations of the receiver's coordinates, $\boldsymbol{r} _i$, the ionospheric delays at the location of the receiver, $\boldsymbol{I} _i$, and the receiver's differential clock offsets, $\Delta \boldsymbol{\tau} _i$, are obtained as their marginal \aposteriori{} distribution: $p(\, \boldsymbol{r}_{(1,2,\ldots,n)} \,|\, \Delta \boldsymbol{N},\, \boldsymbol{y}_{(1,2,\ldots,n)} )$, $p(\, \boldsymbol{I}_{(1,2,\ldots,n)} \,|\, \Delta \boldsymbol{N},\, \boldsymbol{y}_{(1,2,\ldots,n)} )$, and others (see Appendix~II), except for the uncertain common increment in $\boldsymbol{I} _i$ and $\Delta \boldsymbol{\tau} _i$ among the satellites and the epochs. Such uncertainty in $\boldsymbol{I} _i$ and $\Delta \boldsymbol{\tau} _i$ arises from their coupling to $\boldsymbol{N}^0$, which is left to be an unknown integer vector even after the correct integer ambiguity is resolved as already mentioned. The \aposteriori{} estimation of $\boldsymbol{I} _i$ and $\Delta \boldsymbol{\tau} _i$, thus, distributes multimodally.

The accuracy of each kind of the \aposteriori{} estimations on a single-frequency receiver, however, depends largely on the accuracy of the \apriori{} estimation of $\boldsymbol{I} _i$ which was used, which may not be highly accurate except for under the condition of the short baseline length of relative positioning, as further described in Section~V-A.

Consider the ambiguity resolution on the other receivers that reside temporally and spatially adjacent to the multiple-frequency receiver with the integer ambiguity already resolved. Such resolution should be thus helped by using the obtained accurate estimation on $\boldsymbol{I} _i$ as an \apriori{} one in their resolution, based on our assumption of its temporal and spatial constancy. Let us call such an initialization the \textit{non-self-}type of \textit{succeeding initialization}. The use of the accurate \apriori{} estimation of $\boldsymbol{I} _i$ remarkably improves the performance in the ambiguity resolution even on a single-frequency receiver, as also shown in the discussion in Section~V-A.

Furthermore, the receiver, once its integer ambiguity has been resolved, gains the potential ability to resolve the possible future re-initialization of its own integer ambiguity efficiently.
Let us call such an initialization the \textit{self-}type of \textit{succeeding initialization}, which is commonly required just after the discontinuity (\ie{}, the mis-track and re-acquisition) of the carrier signals, known as a cycle slip \cite{LeickA2004, HofmannWellenhoffBetal2001}, and is frequent in the receiver in the mobile environments \cite{KondoK2001}. This beneficial effect arises from the accurate estimation having been obtained on both $\Delta \boldsymbol{\tau} _i$ and $\boldsymbol{I} _i$, based on our assumption of their temporal constancy, as well as the above case. Note that the accurate estimation on the receiver's differential clock offsets, $\Delta \boldsymbol{\tau} _i$, is actually achieved only among accurate carrier-phase measurements, and thus its noticeable effect is actually observed only on a multiple-frequency receiver.

Such estimation and reuse of $\Delta \boldsymbol{\tau} _i$ is interfered by the non-informativeness of the wind-up phase, $\psi_i$, \ie{}, by the condition that the attitude of the receiver's antenna is not measured continuously. In such a case, three or more kinds of carrier-phase measurements are required to achieve the noticeable effectiveness of the self-type of succeeding initialization.

These two distinct, \ie{}, self- and non-self-, types of succeeding initialization correspond to the conditions of informative and non-informative $\Delta \boldsymbol{\tau} _i$, respectively. 
Each is assumed to utilize the respective type of pre-initialization measurements with a certain duration and with their integer ambiguities correctly resolved.
\section{Optimal success and error rates} %

This section describes the optimal ambiguity resolution and its success/error rates as formalized to be a fixed-sample Bayes decision. Its optimality was essentially proved by statistical studies.
A decision, or classification, $\delta$, maps $\boldsymbol{\nu}$ (or the measured sample, $\boldsymbol{y}_{(1,2,\ldots,n)}$, in more general discussion) to a value of decision parameter, $\Delta \boldsymbol{N}$, or avoids a judgment:
\begin{gather}
\delta : \mspace{14mu} \boldsymbol{\nu} \rightarrow \mspace{12mu} \Delta \boldsymbol{N} \mspace{12mu} \text{or} \mspace{12mu} \text{none}
.
\nonumber
\end{gather}
The success, error, and reserving rates arisen from $\delta$ are calculated by integrating the marginal \aposteriori{} (joint) distribution \eqref{marginal_joint_distribution} as follows:
\begin{align}
& \alpha_\delta = \sum_{\Delta \boldsymbol{N}} \int_{D_\delta (\Delta \boldsymbol{N})} p( \Delta \boldsymbol{N},\, \boldsymbol{\nu} ) \, d \boldsymbol{\nu}
\nonumber
, \\
& \beta_\delta = \sum_{\Delta \boldsymbol{N}} \sum_{\substack{\Delta \boldsymbol{N}^\prime\\\neq \Delta \boldsymbol{N}}} \int_{D_\delta (\Delta \boldsymbol{N}^\prime)} p( \Delta \boldsymbol{N},\, \boldsymbol{\nu} ) \, d \boldsymbol{\nu}
\nonumber
, \\
& \gamma_\delta = 1 - \alpha_\delta - \beta_\delta
\nonumber
,
\end{align}
in which $D_\delta (\Delta \boldsymbol{N})$ is defined to be the acceptance region where the decision $\delta$ chooses $\Delta \boldsymbol{N}$ in the space of $\boldsymbol{\nu}$.

These rates are calculated by using \eqref{general_conditional_distribution} and our assumption of the non-informativeness of $p(\boldsymbol{N})$, \apriori{} distribution of $\boldsymbol{N}$ (= $\Delta \boldsymbol{N} + \boldsymbol{N}^0$), as follows:
\begin{align} 
\alpha_\delta & = \int_{D_\delta (\Delta \boldsymbol{N})} p(\,\boldsymbol{\nu} \,|\, \Delta \boldsymbol{N},\, \boldsymbol{N}^0 ) \, d \boldsymbol{\nu}
\nonumber
\\
& = \sqrt{\frac{| M |}{\left( 2 \pi \right)^p}}
\nonumber
\\
& \mspace{28mu} \times
\int_{D_\delta (\Delta \boldsymbol{N})} \negthickspace \exp \negmedspace \left\{ - \frac{1}{2} \left( \boldsymbol{\nu} - \boldsymbol{N} \right)^T \negthinspace M \left( \boldsymbol{\nu} - \boldsymbol{N} \right) \right\} d \boldsymbol{\nu}
,
\label{integration_success_rate}
\end{align}
and
\begin{align}
\beta_\delta & = \sum_{\substack{\Delta \boldsymbol{N}^\prime\\\neq \Delta \boldsymbol{N}}} \int_{D_\delta (\Delta \boldsymbol{N}^\prime)} \, p(\,\boldsymbol{\nu} \,|\, \Delta \boldsymbol{N},\, \boldsymbol{N}^0 ) \, d \boldsymbol{\nu}
\nonumber
\\
& = \sqrt{\frac{| M |}{\left( 2 \pi \right)^p}} \sum_{\substack{\Delta \boldsymbol{N}^\prime\\\neq \Delta \boldsymbol{N}}}
\nonumber
\\
& \qquad
\int_{D_\delta (\Delta \boldsymbol{N}^\prime)} \negthickspace \exp \negmedspace \left\{ - \frac{1}{2} \left( \boldsymbol{\nu} - \boldsymbol{N} \right)^T \negthinspace M \left( \boldsymbol{\nu} - \boldsymbol{N} \right) \right\} d \boldsymbol{\nu}
\label{integration_error_rate_1}
,
\end{align}
where the independence of $\alpha_\delta$ and $\beta_\delta$ on $\Delta \boldsymbol{N}$ and $\boldsymbol{N}^0$ is used. Consequently, the integration domain in \eqref{integration_error_rate_1} is $\bigcup_{\Delta \boldsymbol{N}^\prime \neq \Delta \boldsymbol{N}} D_\delta (\Delta \boldsymbol{N}^\prime)$ with regard to $\boldsymbol{\nu}$.

\putfloat[t]{voronoi polytopes} %
\putfloat[t]{errorintegral} %

A maximum marginal \aposteriori{} decision, $\delta_\mathrm{MAP}$, is defined by using the marginal \aposteriori{} distribution as:
\begin{equation}
\begin{aligned}
& \delta_\mathrm{MAP} : \mspace{14mu} \boldsymbol{\nu} \rightarrow \Delta \boldsymbol{N}_\mathrm{MAP}
,
\\
& \Delta \boldsymbol{N}_\mathrm{MAP} = \underset{\Delta \boldsymbol{N}}{\arg \max} \ p( \Delta \boldsymbol{N} \,|\, \boldsymbol{\nu} )
,
\end{aligned}
\label{maximum_marginal_aposteriori_decision}
\end{equation}
and never avoids a judgment. The decision is well known to maximize $\alpha_\delta$ subject to $\gamma_\delta = 0$, and this optimality was essentially proved by \cite{HoelPGetPetersonRP1949}\footnote{This was rediscovered by \cite{TeunissenPJG1999c} in the field of GPS/GNSS.}. Its \textit{unconditionally} optimal rates, $\alpha_\mathrm{MAP}$ and $\beta_\mathrm{MAP}$, should be calculated by the integrations \eqref{integration_success_rate} and \eqref{integration_error_rate_1} with $\delta_\mathrm{MAP}$ applied. Its acceptance region is denoted as $D_\mathrm{MAP} (\Delta \boldsymbol{N})$.

A \textit{conditionally} optimal decision, $\delta_\mathrm{opt}$, is defined as the modification of $\delta_\mathrm{MAP}$:
\begin{equation}
\delta_\mathrm{opt} : \mspace{14mu} \boldsymbol{\nu} \rightarrow
\begin{cases}
\Delta \boldsymbol{N}_\mathrm{MAP} , & \text{if} \mspace{12mu} p( \Delta \boldsymbol{N}_\mathrm{MAP} \,|\, \boldsymbol{\nu} ) \geq h ; \\
\text{none} , & \text{otherwise} ,
\end{cases}
\label{conditionally_optimal_decision}
\end{equation}
in which a Bayesian confidence parameter, $h \ (\geq \nicefrac{1}{2})$, is introduced, and the case of equality is incorporated with the greater case for the sake of simplicity. $\delta_\mathrm{opt}$ is also well known to maximize $\alpha_\delta$ subject to $\beta_\delta \leq \beta_h$ with $\beta_h$ assigned, and this optimality was essentially proved by \cite{NeymanJetPearsonES1933}\footnote{This was also rediscovered by \cite{TeunissenPJG2004a} in the field of GPS/GNSS.} (see \cite{LehmannEL1986, SchervishMJ1995}).

Its conditionally optimal rates, $\alpha_\mathrm{opt}$, $\beta_\mathrm{opt}$, and $\gamma_\mathrm{opt}$, should be also calculated by the integrations \eqref{integration_success_rate} and \eqref{integration_error_rate_1} by applying $\delta_\mathrm{opt}$ and choosing the value of $h$ so that $\beta_\mathrm{opt}$ should equal $\beta_h$, assuming that $\beta_h$ is assigned to be less than $\beta_\mathrm{MAP}$. Or else, (raw) $\delta_\mathrm{MAP}$ should be used, and then $\alpha_\mathrm{opt} = \alpha_\mathrm{MAP}$, $\beta_\mathrm{opt} = \beta_\mathrm{MAP}$, and $\gamma_\mathrm{opt} = 0$. Let us explicitly denote the dependence on $h$ of $D_\mathrm{opt} (\Delta \boldsymbol{N})$, $\alpha_\mathrm{opt}$, $\beta_\mathrm{opt}$, and $\gamma_\mathrm{opt}$, and omit their subscript, ``opt,'' in the following discussion when they will not be misunderstood, as follows: $D (\Delta \boldsymbol{N}; h)$, $\alpha (h)$, $\beta (h)$, and $\gamma (h)$.

The following approximate equations hold:
\begin{gather}
D (\Delta \boldsymbol{N}; \nicefrac{1}{2}) \simeq \, D_\mathrm{MAP}(\Delta \boldsymbol{N}) ,
\nonumber
\\
\alpha(\nicefrac{1}{2}) \simeq \alpha_\mathrm{MAP} ,
\nonumber
\\
\beta(\nicefrac{1}{2}) \simeq \beta_\mathrm{MAP} .
\nonumber
\end{gather}
The difference between the right-hand and left-hand sides of each approximate equation is extremely small under the condition of the high success rate of $\alpha_\mathrm{MAP}$, or the low error rate of $\beta_\mathrm{MAP}$.

There exists a slight probability that the following inequality holds:
\begin{gather}
p( \Delta \boldsymbol{N}_\mathrm{MAP} \,|\, \boldsymbol{\nu} ) < \nicefrac{1}{2}
 .
\label{slight probability}
\end{gather}
This probability is extremely low under the condition of the extremely high success rate of $\alpha_\mathrm{MAP}$.

Furthermore, let us normalize $h$ and introduce $h^\prime$ so that the maximum of the marginal \aposteriori{} distribution exactly corresponds to $h^\prime = 1$, as follows:
\begin{align}
& h^\prime = \frac{h}{h_0}
,
\nonumber
\\
& h_0 = p( \Delta \boldsymbol{N} \,|\, \boldsymbol{\nu} = \Delta \boldsymbol{N} + \boldsymbol{N}^0 ) \simeq 1
.
\nonumber
\end{align}
In the following discussion, the normalized $h^\prime$ is used instead of the non-normalized $h$. The functions of success, error, and reserving rates will be represented as functions of $h^\prime$, such as $D (\Delta \boldsymbol{N}; h^\prime)$, $\alpha (h^\prime)$, $\beta (h^\prime)$, and $\gamma (h^\prime)$.
\section{The methodologies of the calculations of the rates} %

This section explains the methodologies of the calculations for the optimal success/error rates of integer ambiguity resolution. They are basically a union-bound approximation and a Monte Carlo integration.
\subsection{Union-bound approximation} %

\subsubsection{$\alpha_\mathrm{MAP}$ and $\beta_\mathrm{MAP}$ in the $\Delta \boldsymbol{N}$-separable case} %

Let us firstly consider the calculations of $\alpha_\mathrm{MAP}$ \eqref{integration_success_rate} and $\beta_\mathrm{MAP}$ \eqref{integration_error_rate_1} in the $\Delta \boldsymbol{N}$-separable case. Note that the actual integrand is \eqref{Delta_conditional_distribution} using $M_\Delta$ and the integration variable is $\Delta \boldsymbol{\nu}$ in this case. Generally, if the marginal joint distribution \eqref{Delta_conditional_distribution_0} is a monotonically decreasing function of the positive-definite quadratic form of $\Delta \boldsymbol{N} - \Delta \boldsymbol{\nu}$, the acceptance region, $D_\mathrm{MAP} (\Delta \boldsymbol{N})$, is shaped into a uniform point-symmetrical \textit{Voronoi polytope} in the space of $\Delta \boldsymbol{\nu}$, as shown in Fig.~\ref{voronoi polytope in separable case}. Each lattice point, $\Delta \boldsymbol{N}$, resides at the center of the polytope.

\textit{Voronoi-relevant}, \ie{}, adjacent lattice points, are defined, as their distinct Voronoi polytopes share facets. The adjacent lattice points can be represented as displacement vectors from a certain lattice point to its adjacent ones. Let us denote the set of adjacent integer lattice points as
\begin{equation}
\left\{ \boldsymbol{n}_1,\, \boldsymbol{n}_2,\, \ldots,\, \boldsymbol{n}_i,\, \ldots,\, \boldsymbol{n}_q \right\}
,
\label{adjacent_lattice}
\end{equation}
which contains $q$ lattice points. This set contains all possible lattice points, but any two of them, $\boldsymbol{n}_i$ and $\boldsymbol{n}_j$, should satisfy,
\begin{equation}
\begin{gathered}
\left| {\boldsymbol{n}_i}^T M_\Delta \, \boldsymbol{n}_j \right| < {a_i}^2
,
\\
a_i = \sqrt{{\boldsymbol{n}_i}^T M_\Delta \, \boldsymbol{n}_i}
,
\end{gathered}
\label{generalized_distance}
\end{equation}
in which the generalized distance, $a_i$, of each adjacent lattice point from the origin is introduced. The representation of \eqref{adjacent_lattice} will not be assumed to distinguish between point-symmetrical lattice points, \ie{}, $\boldsymbol{n}_i$ and $-\boldsymbol{n}_i$; thus, they are counted as one lattice point in \eqref{adjacent_lattice}.

Each adjacent lattice point, $\boldsymbol{n}_i$, corresponds to a facet of the Voronoi polytope, which is a generalized perpendicular bisector of $\boldsymbol{n}_i$. At the bisection point, the following approximate equality holds under the high success rate and low error rate conditions:
\begin{equation}
p(\Delta \boldsymbol{N} \: | \: \Delta \boldsymbol{\nu} = \Delta \boldsymbol{N} \pm \nicefrac{1}{2} \; \boldsymbol{n}_i) \simeq \nicefrac{1}{2}.
\nonumber
\end{equation}
The number of facets is $2q$, and $q$ is bounded by the inequality \cite{MinkowskiH1897},
\begin{equation}
p_\Delta \leq q \leq 2^{(p_\Delta)} - 1.
\nonumber
\end{equation}
If $M_\Delta$ is a diagonal matrix, it produces a rectangular Voronoi polytope and $q$ equals $p_\Delta$. For a general matrix, however, $q$ usually takes on its maximum value, $2^{(p_\Delta)} - 1$ \cite{VoronoiG1908}.

It is well known that the union bound \cite{WozencraftJMetJacobsIM1965,ConwayJHetSloaneNJA1999,ProakisJG2000} is the lower one to $\alpha_\mathrm{MAP}$ (and the upper one to $\beta_\mathrm{MAP}$), and the minimum-distance bound \cite{PetersonWWetWeldonJrEJ1972,HammingRW1986} is the upper one to $\alpha_\mathrm{MAP}$ (and the lower one to $\beta_\mathrm{MAP}$), as follows:
\begin{align}
& \alpha_{\textrm{MAP},\min} > \alpha_\textrm{MAP} > 1 - \sum_{i = 1}^{q} \beta_{\textrm{MAP},i} \simeq \prod_{i = 1}^{q} \alpha_{\textrm{MAP},i}
,
\label{alphaMAP}
\\
& \beta_{\textrm{MAP},\min} < \beta_\textrm{MAP} < \sum_{i = 1}^{q} \beta_{\textrm{MAP},i}
,
\label{betaMAP}
\\
& \alpha_{\textrm{MAP},i} = \frac{\;\, 2 a_i}{\sqrt{2 \pi}} \int_{0}^{\nicefrac{1}{2}} \exp \negmedspace \left( -\frac{1}{2} {a_i}^2 x^2 \right) d x
,
\label{alphaMAPi}
\\
& \beta_{\textrm{MAP},i} = \frac{\;\, 2 a_i}{\sqrt{2 \pi}} \int_{\nicefrac{1}{2}}^{\infty} \exp \negmedspace \left( -\frac{1}{2} {a_i}^2 x^2 \right) d x = 1 - \alpha_{\textrm{MAP},i}
,
\label{betaMAPi}
\\
& \alpha_{\textrm{MAP},\min} = \frac{\;\, 2 a_{\min}}{\sqrt{2 \pi}} \int_{0}^{\nicefrac{1}{2}} \exp \negmedspace \left( -\frac{1}{2} {a_{\min}}^2 x^2 \right) d x
,
\nonumber
\\
& \beta_{\textrm{MAP},\min} = \max_i \beta_{\textrm{MAP},i} = 1 - \alpha_{\textrm{MAP},\min}
, 
\nonumber
\\
& a_{\min} = \min_i a_i
.
\nonumber
\end{align}
\subsubsection{$\alpha(h^\prime)$ and $\beta(h^\prime)$ in the $\Delta \boldsymbol{N}$-separable case} %

Secondly, let us consider the calculations of $\alpha(h^\prime)$ and $\beta(h^\prime)$ in the $\Delta \boldsymbol{N}$-separable case. The shape of $D (\Delta \boldsymbol{N}; h^\prime)$ remains as the incomplete Voronoi polytope whose pseudo-facet is tangent to the ellipsoidal equivalence surface of the function \eqref{Delta_conditional_distribution} shown in Fig.~\ref{voronoi polytope in separable case}, based on our assumption of the high success rate and low error rate conditions.

Then, let us introduce one-dimensional \aposteriori{} distributions, each of which will determine the position of the pseudo-facets, as follows:
\begin{gather}
p_i(n\,|\,\xi) = 
\frac{ \exp \negmedspace \left\{ -\frac{1}{2} {a_i}^2 (n - \xi)^2 \right\} }
{ \sum_{ \begin{subarray}{l}n^\prime =\\ -\infty\end{subarray}}^{\infty} \ \exp \negmedspace \left\{ -\frac{1}{2} {a_i}^2 (n^\prime - \xi)^2 \right\} }
.
\nonumber
\end{gather}
Its normalization factor is
\begin{gather}
h_{0,i} = p_i(n\,|\,\xi = n) = \left\{
\sum_{ n^\prime = -\infty }^{\infty} \exp \negmedspace \left( -\frac{1}{2} {a_i}^2 {n^\prime}^2 \right) \right\}^{-1}
\simeq 1
,
\nonumber
\end{gather}
and the value at the bisection point is
\begin{gather}
p_i(n \,|\, \xi = n \pm \nicefrac{1}{2}) \simeq \nicefrac{1}{2}
.
\nonumber
\end{gather}
The last two approximate equalities are derived from the same assumption as mentioned above. Fig.~\ref{errorintegral} shows one-dimensional error distribution, $p_i(\xi \,|\, n = 0)$, one-dimensional \aposteriori{} distribution, $p_i(n\,|\,\xi)$, and the acceptance regions in the space of $\xi$.

We propose the approximated bounds to $\alpha(h^\prime)$ and $\beta(h^\prime)$, which are the variants of the union and minimum-distance ones, as follows:
\begin{align}
& \alpha_{\min}(h^\prime) \gtrsim \alpha(h^\prime) \gtrsim \prod_{i = 1}^{q} \alpha_i(h^\prime)
,
\label{alpha}
\\
& \beta_{\min}(h^\prime) \lesssim \beta(h^\prime) \lesssim \sum_{i = 1}^{q} \beta_i(h^\prime)
,
\label{beta}
\\
& \alpha_i(h^\prime) = \frac{\;\, 2 a_i}{\sqrt{2 \pi}} \int_{0}^{\xi_i} \exp \negmedspace \left( -\frac{1}{2} {a_i}^2 x^2 \right) d x
,
\label{alphai}
\\
& \beta_i(h^\prime) = \frac{\;\, 2 a_i}{\sqrt{2 \pi}} \int_{1-\xi_i}^{\infty} \exp \negmedspace \left( -\frac{1}{2} {a_i}^2 x^2 \right) d x
,
\label{betai}
\\
& \alpha_{\min}(h^\prime) = \frac{\;\, 2 a_{\min}}{\sqrt{2 \pi}} \int_{0}^{\xi_i} \exp \negmedspace \left( -\frac{1}{2} {a_{\min}}^2 x^2 \right) d x
,
\nonumber
\\
& \beta_{\min}(h^\prime) = \max_i \beta_i(h^\prime)
,
\nonumber
\end{align}
where $\xi_i$ should be calculated by using
\begin{equation}
p_i(0 \,|\, \xi_i) = h^\prime \, h_{0,i}
,
\nonumber
\end{equation}
with $h^\prime$ assigned.
These formulas coincide with \eqref{alphaMAP}--\eqref{betaMAPi} by substituting $\xi_i = \nicefrac{1}{2}$. The right-hand side of \eqref{beta} gives the true upper bound to $\beta(h^\prime)$ in the $\Delta \boldsymbol{N}$-separable case, whereas the right-hand side of \eqref{alpha} is expected to give the approximate lower bound to $\alpha(h^\prime)$.

Let us consider the functions of $x$, $\alpha(x)$ and $1 - \beta(1 - x)$, which are approximately bounded by the above formulas. They approximately coincide in the neighborhood of $x = \nicefrac{1}{2}$, because of the following small one-dimensional error rate at $\xi_i= \nicefrac{1}{2}$~:
\begin{equation}
\beta_i(\xi_i = \nicefrac{1}{2}) \, = \, 1 - \alpha_i(\xi_i= \nicefrac{1}{2}) \, \ll \, 1 .
\nonumber
\end{equation}
\subsubsection{Success/error rates in the $\Delta \boldsymbol{N}$-nonseparable case} %

Finally, let us consider the calculations of $\alpha_\mathrm{MAP}$, $\beta_\mathrm{MAP}$, $\alpha(h^\prime)$, and $\beta(h^\prime)$ in the $\Delta \boldsymbol{N}$-nonseparable case. In this case, the marginal \aposteriori{} distribution \eqref{marginal_aposteriori_distribution_1} is expressed by using $M$ (not $M_\Delta$), whose acceptance region in the space of $\boldsymbol{\nu}$ (not $\Delta \boldsymbol{\nu}$) is shaped as shown in Fig.~\ref{voronoi polytope in nonseparable case}. We can define adjacent displacement vectors, $\boldsymbol{n}_i$'s, as well as \eqref{adjacent_lattice}, but by using $M$ instead of $M_\Delta$. Thus, we propose to use the same approximated bounds as \eqref{alphaMAP}--\eqref{betaMAPi} and \eqref{alpha}--\eqref{betai} in this case, with the proviso that we should use such adjacent lattice points, $\boldsymbol{n}_i$'s, that are not any possible vectors of $\boldsymbol{N}^0$'s:
\begin{equation}
\left\{ \boldsymbol{n}_1,\, \boldsymbol{n}_2,\, \ldots,\, \boldsymbol{n}_i,\, \ldots,\, \boldsymbol{n}_{q^\prime} \right\}, \ \nexists \boldsymbol{N}^0 = \boldsymbol{n}_i
\label{adjacent_lattice_2}
.
\end{equation}
Their adjacency condition requires that any two of them should satisfy,
\begin{equation}
\begin{gathered}
\left| {\boldsymbol{n}_i}^T M \, \boldsymbol{n}_j \right| < {a_i}^2
,
\\
a_i = \sqrt{{\boldsymbol{n}_i}^T M \, \boldsymbol{n}_i}
,
\end{gathered}
\label{generalized_distance_2}
\end{equation}
as well as \eqref{generalized_distance}.

This set exactly provides the set \eqref{adjacent_lattice} derived from $M_\Delta$, if we transform $M$ to a $\Delta \boldsymbol{N}$-separable matrix. (Consider, for example, the transformation of the condition of Fig.~\ref{voronoi polytope in nonseparable case} to that of Fig.~\ref{errorintegral}.)
\subsubsection{Modifications} %

Although the approximated union bounds being proposed are expected to be tight under the high success rate and low error rate conditions, their tightness is degraded by any condition that lowers the success rate $\alpha(h^\prime)$ by increasing $h^\prime$. The approximated minimum-distance bound proposed for $\beta(h^\prime)$ may, furthermore, fail to bound the exact value under such conditions, especially in the $\Delta \boldsymbol{N}$-nonseparable case.

In order to improve their accuracy under such conditions, we furthermore propose to modify both the bound to error rate \eqref{beta} and the one-dimensional error rate \eqref{betai} into 
\begin{align}
& \beta_{\min}(h^\prime) \prod_{\substack{i = 1\\( i \neq \min)}}^{q \, \text{or}\, q^\prime} \alpha_i(h^\prime) \, \lesssim \, \beta(h^\prime) \, \lesssim \, \sum_{i=1}^{q \, \text{or}\, q^\prime} \beta_i(h^\prime)
,
\label{beta2}
\\
& \beta_i(h^\prime) = \frac{\;\, 2 a_i}{\sqrt{2 \pi}} \int_{1-\xi_i}^{1+\xi_i} \negthickspace \exp \negmedspace \left( -\frac{1}{2} {a_i}^2 x^2 \right) d x
,
\label{betai2}
\\
& \beta_{\min}(h^\prime) = \max_i \beta_i(h^\prime)
.
\nonumber
\end{align}
These modifications, \eqref{beta2} and \eqref{betai2}, will be used instead of \eqref{beta} and \eqref{betai} in the following calculations. In general, an excess degeneracy of the polytope's facets closer to the center point, or a large number of the adjacent lattice points with small $a_i$, may degrade the accuracy of the union and minimum-distance bounds.

If the single approximation formulas are preferred instead of the bound ones, we propose to use the following approximation for $\beta(h^\prime)$:
\begin{equation}
\beta(h^\prime) \sim \sum_{i=1}^{q \, \text{or}\, q^\prime} \beta_i(h^\prime) \prod_{\substack{j=1\\(j \neq i)}}^{q \, \text{or}\, q^\prime} \alpha_j(h^\prime) 
,
\label{beta3}
\end{equation}
and the lower bound in \eqref{alpha} for $\alpha(h^\prime)$.
\subsection{Importance-sampling Monte Carlo integration} %

The approximated union bound proposed in Section~IV-A is not always tight enough to investigate the rigorous value of $\alpha_\mathrm{MAP}$, $\beta_\mathrm{MAP}$, $\alpha(h^\prime)$, and $\beta(h^\prime)$, even under the high success rate and low error rate conditions. A numerical Monte Carlo method \cite{HammersleyJMetHandscombDC1964}, by contrast, always provides the reliable calculation of the rates, but suffers from the severe slowness of numerical convergence in high-dimensional integrations. This study will accelerate the convergence by using the importance sampling method \cite{HammersleyJMetHandscombDC1964} described below.

\subsubsection{Basic formalization} %

Consider the integration of multivariate Gaussian distribution inside a certain closed domain, $D$, as follows:
\begin{equation}
I = \sqrt{\frac{|M|}{(2 \pi)^p}} \int_{D} \exp \negmedspace \left( -\frac{1}{2} \boldsymbol{x}^T M \: \boldsymbol{x} \right) d \boldsymbol{x}
,
\label{Monte_Carlo_integration_1}
\end{equation}
where $p$ is the size of the inverse covariance matrix, $M$. $I$ is assumed to be $\alpha_\delta$ in \eqref{integration_success_rate} or $\beta_\delta$ in \eqref{integration_error_rate_1} with $\boldsymbol{x}$ set to $\boldsymbol{\nu} - \boldsymbol{N}$ and $D$ is, thus, assumed to be $D_\delta(\Delta \boldsymbol{N} \negthinspace = \negthinspace \boldsymbol{0})$ or $\bigcup_{\Delta \boldsymbol{N}^\prime \neq \boldsymbol{0}} D_\delta (\Delta \boldsymbol{N}^\prime)$, respectively, with regard to $\boldsymbol{x}$.

This study also assumes that random points, $\boldsymbol{\xi}_i$'s, are generated for a Monte Carlo integration for $I$ with regard to $\boldsymbol{x}$ and distribute according to distinct multivariate Gaussian distribution with the inverse covariance matrix, $\tilde{M}$:
\begin{equation}
\left\{ \boldsymbol{\xi}_i \right\} \underset{\text{distrib.}}{\thicksim} p(\boldsymbol{\xi}) = \sqrt{\frac{|\tilde{M}|}{(2 \pi)^p}} \exp \negmedspace \left( -\frac{1}{2} \boldsymbol{\xi}^T \tilde{M} \: \boldsymbol{\xi} \right)
.
\label{importance_sampling_1}
\end{equation}
The hit-or-miss Monte Carlo integration \cite{HammersleyJMetHandscombDC1964} for $I$ is, thus, described as:
\begin{equation}
\tilde{I}_N = \frac{1}{N} \sum_{i=1}^N \sqrt{\frac{|M|}{|\tilde{M}|}} \exp \negmedspace \left\{ -\frac{1}{2} {\boldsymbol{\xi}_i}^T \negthinspace \left( M - \tilde{M} \right) \boldsymbol{\xi}_i \right\} \varDelta(\,\boldsymbol{\xi}_i \in D)
,
\nonumber
\end{equation}
where $N$ is the total number of generations and a ``0-1'' reward function is used: $\varDelta(\mathcal{P})$ = 1, if $\mathcal{P}$ is true; 0, otherwise.
This study uses the reliable algorithm \cite{MatsumotoMetNishimuraT1998} for the generation of random numbers.

The coefficient of the square of the error variance, $\kappa$, is defined and calculated as follows:
\begin{align}
\kappa \; & = \lim_{N \rightarrow \infty} \left\langle \left( \tilde{I}_N - I \right)^2 \right\rangle \times N
\nonumber
\\
& = \sqrt{\frac{|M|^2}{(2 \pi)^p |\tilde{M}|}} \int_{D} \exp \negmedspace \left\{ -\frac{1}{2} \boldsymbol{x}^T \negthinspace \left( 2 M - \tilde{M} \right) \boldsymbol{x} \right\} d\boldsymbol{x}
\label{kappa}
,
\end{align}
where we used the assumption that the number of the drops of $\boldsymbol{\xi}_i$'s in $D$ is much less than $N$. This assumption is justified if we consider $\beta_\delta$ as $I$ under the high success rate and low error rate conditions.

We numerically optimize each element of $\tilde{M}$ to minimize $\kappa$ for $\beta(h^\prime)$ by using the Broyden-Fletcher-Goldfarb-Shanno method \cite{DennisJEetSchnabelRB1983, FletcherR1987}, where the integration part in \eqref{kappa} is normalized as:
\begin{equation}
\sqrt{\frac{|2 M - \tilde{M}| }{(2 \pi)^p }} \int_{D} \exp \negmedspace \left\{ -\frac{1}{2} \boldsymbol{x}^T \negthinspace \left( 2 M - \tilde{M} \right) \boldsymbol{x} \right\} d\boldsymbol{x}
\nonumber
,
\end{equation}
and is calculated by using \eqref{beta3}, in which $\alpha_i(h^\prime)$ and $\beta_i(h^\prime)$, \ie{}, \eqref{alphai} and \eqref{betai2}, should use distinct $a_i^\prime$ instead of $a_i$ in \eqref{generalized_distance_2} as follows:
\begin{equation}
a_i^\prime = \frac{{\boldsymbol{n}_i}^T M \, \boldsymbol{n}_i}{ \sqrt{ \boldsymbol{n}_i^T M \left( 2 M - \tilde{M} \right)^{-1} \negmedspace M \, \boldsymbol{n}_i } }
\nonumber
.
\end{equation}

\subsubsection{Another type of formalization} %

Another type of the distribution for importance sampling accelerates the Monte Carlo integration for $\beta(h^\prime)$ with large $h^\prime$. It is derived from the following modification of the integration:
\begin{align}
& \beta(h^\prime) = \sqrt{\frac{| M |}{\left( 2 \pi \right)^p}} \sum_{ \Delta \boldsymbol{N}^\prime \neq \Delta \boldsymbol{N} }
\nonumber
\\
& \qquad
\int_{D(\Delta \boldsymbol{N} ;\, h^\prime)} \negthickspace \exp \negmedspace \left\{ - \frac{1}{2} \left( \boldsymbol{x} - \boldsymbol{N}^\prime \right)^T \negthinspace M \left( \boldsymbol{x} - \boldsymbol{N}^\prime \right) \right\} d \boldsymbol{x}
,
\label{Monte_Carlo_integration_2}
\end{align}
where $\boldsymbol{x}$ is assumed to be $\boldsymbol{\nu} - \boldsymbol{N} + \boldsymbol{N}^\prime$ in \eqref{integration_error_rate_1}, and thus the integration domain is $D(\Delta \boldsymbol{N} ;\, h^\prime)$ with regard to $\boldsymbol{x}$. We also assumes random points, $\boldsymbol{\xi}_i$'s, are generated for $\boldsymbol{x}$ and distribute according to the same form of multivariate Gaussian distribution with the inverse covariance matrix, $\tilde{M}$, as \eqref{importance_sampling_1}.This type of the importance-sampling Monte Carlo integration for \eqref{Monte_Carlo_integration_2} with optimized $\tilde{M}$ provides the faster convergence for the integration of $\beta(h^\prime)$ in the region of about $\log_{10} h^\prime / ( 1 - h^\prime ) > 15$ in the calculations of Section~IV-D, whereas the former type of the optimized Monte Carlo integration for \eqref{Monte_Carlo_integration_1} suffers from the slowness of the convergence in that region.
\subsection{Efficient calculations of the optimal decisions} %

This subsection explains how to improve the efficiency in the critical parts of the calculations for the optimal decisions in this study. The first half introduces the basics of the algorithms and the last half proposes their enhancements required in this study.
\subsubsection{Sphere decoding} %

We use the efficient algorithm for the search and the enumeration in an integer lattice involved in the optimal decisions, which was essentially developed by Pohst and Fincke \cite{PohstM1981,FinckeUetPohstM1985}, and is known as a sphere decoding, and has been applied to such problems as a closest vector one, a shortest vector one \cite{CohenH1993}, and a communication decoder \cite{ViterboEetBiglieriE1993, MowWH1994}. It is the branch-and-bound-type search \cite{HorowitzEetSahniS1978,AhoAVetal1983} down to the bottom layer of the search tree, which enumerates the lattice points, $\boldsymbol{N}$, enclosed by a $p$-dimensional base ellipsoid. This corresponds to the inequality for the quadratic form using $\boldsymbol{\nu}$ and $M$ in \eqref{general_conditional_distribution}, as follows:
\begin{equation}
\begin{aligned}
& \left( \boldsymbol{N} - \boldsymbol{\nu} \right)^T M \left( \boldsymbol{N} - \boldsymbol{\nu} \right) \\
=& \left( \boldsymbol{N}^\prime - \boldsymbol{\nu}^\prime \right)^T {L^\prime}^T L^\prime \left( \boldsymbol{N}^\prime - \boldsymbol{\nu}^\prime \right) \leq \chi^2 ,
\label{base_inequality}
\end{aligned}
\end{equation}
where
\begin{equation}
\begin{aligned}
{L^\prime}^T L^\prime &= M^\prime = \left( Z^{-1} \right)^T \negthinspace M Z^{-1} ,\\
\boldsymbol{N}^\prime &= Z \boldsymbol{N} = \begin{pmatrix} N_1^\prime,\, N_2^\prime,\, \ldots,\, N_p^\prime \end{pmatrix}^T,\\
\boldsymbol{\nu}^\prime &= Z \boldsymbol{\nu} = \begin{pmatrix} \nu_1^\prime,\, \nu_2^\prime,\, \ldots,\, \nu_p^\prime \end{pmatrix}^T,
\end{aligned}
\nonumber
\end{equation}
and the ``radius,'' $\chi$, of the base ellipsoid is introduced. $M$ is preprocessed to ${\smash[b]{\mathstrut M}}^\prime$ by using the transformation with an integer unimodular matrix, $Z$, in order to reduce the lattice basis ahead of the actual search as further described in the paragraph after next. The lower triangular matrix, ${\smash[b]{\mathstrut L}}^\prime$, is calculated by using the Cholesky decomposition of ${\smash[b]{\mathstrut M}}^\prime$.

The search is comprised of the self-homogeneous searches for the sub-dimensional lattice points enclosed also by an ellipsoid. The base inequality \eqref{base_inequality} is, consequently, decomposed into the following sequential inequalities:
\begin{equation}
\begin{aligned}
& \left( \boldsymbol{N}_{(1)}^\prime - \boldsymbol{\nu}_{(1)}^\prime \right)^T {L_{(1)}^\prime \negthickspace}^T L_{(1)}^\prime \left( \boldsymbol{N}_{(1)}^\prime - \boldsymbol{\nu}_{(1)}^\prime \right) \\
\leq & \left( \boldsymbol{N}_{(2)}^\prime - \boldsymbol{\nu}_{(2)}^\prime \right)^T {L_{(2)}^\prime \negthickspace}^T L_{(2)}^\prime \left( \boldsymbol{N}_{(2)}^\prime - \boldsymbol{\nu}_{(2)}^\prime \right) \\
 & \phantom{\smash{\left( \boldsymbol{N}_{(2)}^\prime - \boldsymbol{\nu}_{(2)}^\prime \right)^T {L_{(2)}^\prime \negthickspace}^T}}
\vdots \\
\leq & \left( \boldsymbol{N}_{(p)}^\prime - \boldsymbol{\nu}_{(p)}^\prime \right)^T {L_{(p)}^\prime \negthickspace}^T L_{(p)}^\prime \left( \boldsymbol{N}_{(p)}^\prime - \boldsymbol{\nu}_{(p)}^\prime \right) \\
=& \left( \boldsymbol{N}^\prime - \boldsymbol{\nu}^\prime \right)^T {L^\prime}^T L^\prime \left( \boldsymbol{N}^\prime - \boldsymbol{\nu}^\prime \right) \leq \chi^2
,
\end{aligned}
\label{sequential_inequalities}
\end{equation}
in which we have introduced subvectors, $\boldsymbol{N}_{(i)}^\prime = \begin{pmatrix} N_1^\prime,\, N_2^\prime,\, \ldots,\, N_i^\prime \end{pmatrix}^T$, and $\boldsymbol{\nu}_{(i)}^\prime = \begin{pmatrix} \nu_1^\prime,\, \nu_2^\prime,\, \ldots,\, \nu_i^\prime \end{pmatrix}^T$, and a submatrix,
\begin{equation}
L_{(i)}^\prime =
\begin{pmatrix}
L_{11}^\prime & 0 & \cdots & 0 \\
L_{21}^\prime & L_{22}^\prime & \cdots & 0 \\
\vdots & \vdots & \ddots & \vdots \\
L_{i1}^\prime & L_{i2}^\prime & \cdots & L_{ii}^\prime 
\end{pmatrix}
,
\nonumber
\end{equation}
where $i = 1,\, 2,\, \ldots,\, p$.

In each layer in the downward process of the search, any of the incomplete (or complete if $i = p$) solutions, $\boldsymbol{N}_{(i)}^\prime$'s, should satisfy the following inequality:
\begin{equation}
\begin{gathered}
\left( \boldsymbol{N}_{(i)}^\prime - \boldsymbol{\nu}_{(i)}^\prime \right)^T {L_{(i)}^\prime \negthickspace}^T L_{(i)}^\prime \left( \boldsymbol{N}_{(i)}^\prime - \boldsymbol{\nu}_{(i)}^\prime \right) \leq \chi^2
.
\end{gathered}
\nonumber
\end{equation}
The number of the solutions, $\boldsymbol{N}_{(i)}^\prime$'s, is, consequently, roughly estimated to be
\begin{equation}
\eta_i = \frac{\pi^{\nicefrac{i}{2}}}{\Gamma\left(\nicefrac{i}{2}+1\right)} \; \chi^i \prod_{j=1}^i \left(L_{jj}^\prime \right)^{-1} .
\nonumber
\end{equation}
The numerical complexity of the search in the full, or $p$-, dimensional space is, thus, estimated to be
\begin{equation}
\sum_{i=1}^p \eta_i \mspace{14mu} \text{or} \mspace{14mu} \sum_{i=0}^{p-1} \eta_i
.
\label{estimated_complexity}
\end{equation}

The preprocessing reduction of the lattice basis deforms the base ellipsoid in order to diminish such numerical complexity by applying the transformation by $Z$. The celebrated algorithm of the lattice basis reduction introduced by Lenstra, Lenstra, and Lov\'{a}sz \cite{LenstraAKetal1982, CohenH1993} efficiently provides the reasonably descendingly-reordered\footnote{Actually, the original formalization by \cite{LenstraAKetal1982} reorders ascendingly the diagonal elements of the upper triangular matrix derived by the Cholesky decomposition, $M^\prime = {U^\prime}^T U^\prime$, which is commonly used by the implementations of sphere decoding such as \cite{FinckeUetPohstM1985, CohenH1993}.} diagonal elements, ${\smash[b]{\mathstrut L}}^\prime_{j j}$'s, and is widely used in the implementation of sphere decoding as well as in the original one by \cite{FinckeUetPohstM1985}, and is also used in the calculations in Section~V-A.

A suboptimal solution, $\boldsymbol{N}_\mathrm{B}^\prime = \begin{pmatrix} N_{\mathrm{B},1}^\prime,\, N_{\mathrm{B},2}^\prime,\, \ldots,\, N_{\mathrm{B},p}^\prime \end{pmatrix}^T$, is defined as each of its elements being derived by minimizing each side of the inequalities \eqref{sequential_inequalities} sequentially down to the bottom layer. This solution was introduced by \cite{BabaiL1986} and is known as a Babai nearest plane solution\footnote{It is also known as a sequential-rounding resolution, which has been rediscovered and used in the field of GPS/GNSS \cite{BlewittG1989,DongDetBockY1989,TeunissenPJG1998}.}, whose complexity of calculation belongs to the polynomial-time class with regard to $p$. This solution itself is statistically inadmissible\footnote{In \cite{TeunissenPJG1999c}, the admissibility of statistical decisions or classifications was misdefined in the discussion about several kinds of truly invariant integer ambiguity resolutions.} and is not used to calculate the success/error rates in this study. It is used to bootstrap the efficient algorithm described below.
\subsubsection{The Schnorr-Euchner strategy with the dynamic shrink of the radius} %

The search algorithm is furthermore improved by using the following refinements developed by \cite{ViterboEetBiglieriE1993,SchnorrCPetEuchnerM1994, MowWH1994,ChanAMetLeeI2002,AgrellEetal2002}, which are also used in this study. They use the depth-first movement \cite{HorowitzEetSahniS1978,AhoAVetal1983} in the search tree, observe the distancing order with $\boldsymbol{N}_\mathrm{B}^\prime$ as starting point for the branches in each layer of the search tree \cite{SchnorrCPetEuchnerM1994}:
\begin{equation}
\begin{aligned}
& N_i^\prime = N_{\mathrm{B},i}^\prime,\, N_{\mathrm{B},i}^\prime +1,\, N_{\mathrm{B},i}^\prime -1,\, N_{\mathrm{B},i}^\prime +2,\, \ldots , \\
\text{or} \mspace{14mu} & N_i^\prime = N_{\mathrm{B},i}^\prime,\, N_{\mathrm{B},i}^\prime -1,\, N_{\mathrm{B},i}^\prime +1,\, N_{\mathrm{B},i}^\prime -2,\, \ldots ,
\end{aligned}
\nonumber
\end{equation}
and dynamically and rapidly shrink the radius, $\chi$, each time the movement succeeds in reaching a complete solution, $\boldsymbol{N}^\prime$, residing in the bottom layer in the process of the search.

For a closest vector problem or a shortest vector problem, such update of $\chi$ is calculated as follows:
\begin{equation}
\chi^2 = \left( \boldsymbol{N}^\prime - \boldsymbol{\nu}^\prime \right)^T {L^\prime}^T L^\prime \left( \boldsymbol{N}^\prime - \boldsymbol{\nu}^\prime \right) ,
\label{radius_updated}
\end{equation}
which is also used in the calculation for the maximum marginal \aposteriori{} decision \eqref{maximum_marginal_aposteriori_decision} in the $\Delta \boldsymbol{N}$-separable case.
$\boldsymbol{N}_\mathrm{B}^\prime$ is, consequently, the first reached complete solution in this refined algorithm. It actually defines the initial radius, $\chi_\mathrm{B}$, according to \eqref{radius_updated}, and, thus, determines the actual numerical complexity of the search based on the estimation \eqref{estimated_complexity} with $\chi_\mathrm{B}$ substituted in $\eta_i$.


These algorithms are used in this study to make the maximum marginal \aposteriori{} decision \eqref{maximum_marginal_aposteriori_decision} and the conditionally optimal decision \eqref{conditionally_optimal_decision}, both of which are included in the laborious Monte Carlo calculations for the optimal success and error rates. These algorithms are also used to construct the set of Voronoi-relevant lattice points, \eqref{adjacent_lattice} or \eqref{adjacent_lattice_2}. Both types of calculations are known to belong to the nondeterministic-polynomial-time class with regard to $p$. The latter calculation furthermore uses the algorithm introduced by \cite{AgrellEetal2002}, which efficiently examines Voronoi-relevant lattice points, $\boldsymbol{n}_i$'s, by using the following equation:
\begin{equation}
\begin{aligned}
& \left\{ \boldsymbol{N} \in \boldsymbol{Z}^p; \, \left( \boldsymbol{N}- \nicefrac{1}{2} \, \boldsymbol{n}_i \right)^T \negthinspace M \left( \boldsymbol{N}- \nicefrac{1}{2} \, \boldsymbol{n}_i \right) \leq \nicefrac{1}{4} \, {a_i}^2 \right\}
\\
&= \left\{ \boldsymbol{0},\, \boldsymbol{n}_i \right\}
.
\end{aligned}
\nonumber
\end{equation}

In order to reduce the lattice basis by using $Z$ in the step of the preprocessing, we use the Lenstra-Lenstra-Lov\'{a}sz's algorithm with its $\delta$ parameter set to one\footnote{In \cite{TeunissenPJG1995}, the same condition was used accompanied by both rediscoveries of this algorithm for lattice basis reduction and the Fincke-Pohst algorithm for a closest vector problem. In the field of GPS/GNSS, the most methods (see \cite{HassibiAetBoydS1998,AbdelHafezMFetal2002,GrafarendEWP2000}) oriented to the efficient calculation of the integer ambiguity resolution indeed use these algorithms inspired by \cite{TeunissenPJG1995}.}, whose condition has not yet been proved to belong to the polynomial-time class with regard to $p$, unlike the conditions of $1/4 < \delta < 1$. The reordering procedure built in the algorithm of lattice basis reduction should be restricted inside each subspace of $\Delta \boldsymbol{N}$ and $\boldsymbol{N}^0$ in the $\Delta \boldsymbol{N}$-nonseparable case because of the necessity to calculate the marginal joint probability \eqref{marginal_joint_distribution} in the $p_\Delta$-th layer in the process of the search.

The extra column-reordering introduced by \cite{FinckeUetPohstM1985} is not used in this study because it is examined and does not reduce the estimation of the numerical complexity \eqref{estimated_complexity}. The strategy of the lattice basis reduction applied to $M^{-1}$ instead of $M$ also introduced by \cite{FinckeUetPohstM1985} is not used either for the same reason.

\putfloat[t]{conditional rates in separable case} %
\putfloat[t]{conditional rates in nonseparable case} %

This study efficiently calculates the infinite sum appearing at the denominator (and numerator) in the marginal \aposteriori{} distributions, \eqref{marginal_aposteriori_distribution_1} and \eqref{marginal_aposteriori_distribution_2}, by using the finite, \ie{}, approximated, set of $\boldsymbol{N}$'s. This set excludes such $\boldsymbol{N}$'s that only contribute extremely slightly to the sum, and is actually constructed by using the sphere-decoding search with its bounding radius, $\chi$, dynamically updated as follows:
\begin{equation}
\chi^2 = \left( \boldsymbol{N}- \boldsymbol{\nu}\right)^T \negthinspace M \left( \boldsymbol{N}- \boldsymbol{\nu} \right) + c
,
\nonumber
\end{equation}
where a constant, $c$, is added by this study. This approximation is justified when we choose a sufficiently large value for $c$ enough to neglect the truncation error induced.

The similar approximation also improves the efficiency in the calculations of the union bound, \eqref{adjacent_lattice}--\eqref{betaMAPi}, \eqref{alpha}--\eqref{betai}, and \eqref{adjacent_lattice_2}--\eqref{betai2}, or that of the Voronoi-relevant lattice points, by using the approximated set of $a_i$'s which is produced by eliminating such $a_i$'s as larger than given $a_\text{th}$. This approximation is justified as well by the fact of the extremely slight contribution of the union of such large $a_i$'s to the total sum or product for the union bound, when we choose a sufficiently large value for $a_{\text{th}}$ compared with $a_{\min}$. This is also actually realized by using the sphere-decoding search with $\boldsymbol{\nu}$ set to $\boldsymbol{0}$ and its bounding radius, $\chi$ (= $a_\text{th}$), dynamically updated each time a new Voronoi-relevant lattice point, $\boldsymbol{n}_i$, is found in the process of the search, as follows:
\begin{equation}
\chi^2 = {\boldsymbol{n}_i}^T M \boldsymbol{n}_i + c ,
\nonumber
\end{equation}
where a constant, $c$, should be also chosen to be sufficiently large.
\subsection{Calculation test} %

We calculate the accurate success/error rates by using the Monte Carlo integration and examine the tightness of the proposed approximated bounds by comparing them.

$\alpha(h^\prime)$ and $\beta(h^\prime)$ are accurately calculated by the accelerated Monte Carlo integration under two distinct conditions, which are shown in Fig.~\ref{conditional rates in separable case} and \ref{conditional rates in nonseparable case}, respectively, accompanied by those calculated by the approximated upper and lower bounds \eqref{alpha}, \eqref{alphai}, \eqref{beta2}, and \eqref{betai2}. The former condition is chosen in the $\Delta \boldsymbol{N}$-separable case and the latter in the $\Delta \boldsymbol{N}$-nonseparable case. The details of both conditions are carefully chosen based on preliminary surveys and they have comparably high $\alpha_\mathrm{MAP}$, \ie{}, $\log_{10} \alpha_\mathrm{MAP} / ( 1 - \alpha_\mathrm{MAP} ) \approx$ 6.5.

In Fig.~\ref{conditional rates in separable case} and \ref{conditional rates in nonseparable case}, the (exact) $\alpha_\mathrm{MAP}$ and $\beta_\mathrm{MAP}$ are calculated by the Monte Carlo integration and shown as the lowest ends, or the left ends, of the plots in the region of $\log_{10} h^\prime / ( 1 - h^\prime ) < 0$, or $p(\Delta \boldsymbol{N}_\mathrm{MAP} |\,\boldsymbol{\nu}) < \nicefrac{1}{2}$. The almost horizontal section of each plot in this region indicates its slight probability of occurrences, as mentioned in \eqref{slight probability}. Note that $\alpha_\mathrm{MAP} \approx \alpha(1/2)$, $\beta_\mathrm{MAP} \approx \beta(1/2)$, and $\alpha_\mathrm{MAP} = 1 - \beta_\mathrm{MAP}$. The plots of $\alpha(h^\prime)$ and $\beta(h^\prime)$ should be looked at in the region of $\log_{10} h^\prime / ( 1 - h^\prime ) \geq 0$ to discuss the tightness of the bounds in the following discussion.

\putfloat[t]{dependence of success rates} %

\putfloat[t]{distinct_sets_measurements} %
\putfloat[t]{parameters_ Individual} %

The results qualitatively examine the tightness of the bounds to $\alpha_\mathrm{MAP}$, which provide $\log_{10} \alpha_\mathrm{MAP} / ( 1 - \alpha_\mathrm{MAP} ) =$ 6.5008, 6.5007, and 6.5525 by the Monte Carlo integration, the union bound, and the minimum-distance bound, respectively, under the former condition, and 6.5410, 6.4319, and 6.7231 under the latter condition.

The tightness of $\alpha(h^\prime)$ and $\beta(h^\prime)$ in the region near $\log_{10} h^\prime / ( 1 - h^\prime ) = 0$ is almost similar to that of $\alpha_\mathrm{MAP}$ and $\beta_\mathrm{MAP}$ under each condition. The approximated union bound under the former condition is extremely tight in the region of $\log_{10} h^\prime / ( 1 - h^\prime ) \leq 15$. This is regarded as common in this case under the condition of high $\alpha_\mathrm{MAP}$. Under the latter condition, however, the approximated union bound, as well as the minimum-distance bound, is not extremely tight even in the region near $\log_{10} h^\prime / ( 1 - h^\prime ) = 0$. We thus conclude the (approximated) union bound does not always provide its excellent tightness in the $\Delta \boldsymbol{N}$-nonseparable case even under the condition of high $\alpha_\mathrm{MAP}$, and it should not be used as the approximation of the rates in such a case if we want to analyze the performance of the integer ambiguity resolution rigorously.

\section{Applications to the analysis of the integer ambiguity resolution} %

This section analyzes how to realize the high reliability of the short-term integer ambiguity resolution in carrier-phase GPS/GNSS positioning by calculating the optimal success rate, $\alpha_\mathrm{MAP}$, and its dependence on such conditions as the uncertainty in ionospheric delay, the initialization time, and the difference between the static and kinematic resolutions.

\subsection{The dependence on ionospheric delay and initialization conditions} %

\subsubsection{Calculation of $\alpha_\mathrm{MAP}$} %

Fig.~\ref{dependence of success rates} shows the dependence of $\alpha_\mathrm{MAP}$ on the error variance of a single-differentiated ionospheric delay, $\sigma_{\varDelta I}$. Their upper abscissa indicates the baseline length which is simply estimated to be $\sigma_{\varDelta I}$ multiplied by $10^6$. Table~I shows the five distinct sets of code-pseudorange and carrier-phase measurements used in the calculations. Note that the set of $\text{L}_\text{W}$ or $\text{L}_\text{W}$+$\text{L}_\text{EW}$ artificially excludes the inherent equation of L1 carrier-phase measurement from the set in \eqref{wide_lanes}. The precise positioning should be achieved by resolving the integer ambiguities involved in L1+L2 or L1+L2+L5 measurements, or by resolving those in any sets of measurements under the condition of $\sigma_{\varDelta I} \sim 0$ as further discussed later.

The calculations also assume two distinct kinds of conditions (a) and (b), which have a difference in error in the code-pseudorange measurements, and correspond to Figs.~\ref{dependence of success rates a} and \ref{dependence of success rates b}, respectively. The distinct part of these conditions is described in Table~II, and the common part is described in Table~III.
The calculation is executed under the condition of \apriori{} non-time-correlative receiver's coordinates, \ie{}, kinematic (on-the-fly) resolutions.

The variances of the two kinds of error components in measurements, \ie{}, a temporally constant one and a temporally varying one, are always set at the same value as each other, as in Table~II. This means the magnitude of the total error variance is equal to that of the individual one multiplied by $\sqrt{2}$.

\putfloat[t]{parameters_common} %
\putfloat[t]{range-error dependence} %

Fig.~\ref{dependence of success rates a} furthermore contains the plots calculated under the condition of self- and non-self-succeeding initialization using the measurements of L1, L1+L2, or L1+L2+L5. Their measurement time for the ambiguity resolution is assumed to follow the pre-measurement one without interval. The condition of self-succeeding initialization is also assumed to be informed of the accurate values of the phase wind-up by measuring the attitude of the receiver's antenna continuously. Otherwise the success rates is examined to be degraded to the value close to that under the condition of a non-self-succeeding initialization.

Each plot of $\alpha_\mathrm{MAP}$ is essentially calculated by using the Monte Carlo integration. The series of the dense points of the union-bound solutions are also calculated and overwritten on any sections of the plots where the difference between these two kinds of solutions is assessed as extremely slight or undetectable. The points of the Monte Carlo solution are not densely plotted in such sections in order to save total computation time, which is justified by the reasonable assumption of the intrinsic smoothness of the plots.

The whole of a plot, in contrast, is calculated only by using the Monte Carlo integration in the cases of the small value of $\alpha_\mathrm{MAP}$ (\ie{}, $\log_{10} \alpha_\mathrm{MAP} / ( 1 - \alpha_\mathrm{MAP} ) \leq 2$) and in the case of the non-self-succeeding initialization using the L1+L2 measurements, where the union bound is not tight to the exact value and should not be used to plot any of the points. It consumes huge computation time to calculate the whole of a smooth plot, \ie{}, the accurate series of the dense points, by using the Monte Carlo integration, if it has the high value of $\alpha_\mathrm{MAP}$ (\ie{}, $\log_{10} \alpha_\mathrm{MAP} / ( 1 - \alpha_\mathrm{MAP} ) \geq 4$).\subsubsection{Results and considerations} %

Figs.~\ref{dependence of success rates a} and \ref{dependence of success rates b} indicate that the high success rates are achieved at $\sigma_{\varDelta I} = 0$, \ie{}, under the condition of a vanishing baseline length, especially by using the L1+L2+L5 measurements. The use of the L1+L2 measurements provides the high success rate only under the condition of a vanishing baseline length and small error in code-pseudorange measurements shown in Fig.~\ref{dependence of success rates a}. The large error in code-pseudorange measurements noticeably degrades the success rates except for in the case of L1+L2+L5 or $\text{L}_\text{W}$+$\text{L}_\text{EW}$ measurements, shown by comparing Figs.~\ref{dependence of success rates a} and \ref{dependence of success rates b}.

The increase in $\sigma_{\varDelta I} $ severely degrades the success rate especially under the condition of the use of L1+L2+L5 or L1+L2 measurements, which contrasts with that of $\text{L}_\text{W}$ or $\text{L}_\text{W}$+$\text{L}_\text{EW}$ measurements. This proves that the precise positioning essentially suffers from the low reliability of the integer ambiguity resolution under the condition of a long baseline length. The $\text{L}_\text{W}$ or $\text{L}_\text{W}$+$\text{L}_\text{EW}$ measurements have the disadvantage in the precision in positioning under this condition as further discussed below.

The use of the succeeding initialization almost overcomes the \apriori{} uncertainty in ionospheric delay by actually obtaining the accurate estimation of $\sigma_{\varDelta I}$, which is derived from the use of the pre-initialization measurements with their ambiguities correctly resolved. The self-succeeding initialization furthermore almost overcomes the \apriori{} uncertainty in the differential clock offset of the receiver and gains the increase of the success rate, shown by comparing it with the case of non-self-succeeding initialization.
\subsubsection{Precision in positioning} %

Fig.~\ref{dependence of range error} shows the dependence of the variance of range errors on that of the error in single-differentiated ionospheric delay for each kind of measurements being analyzed in this study. Each range-error variance corresponds to the least measurement error achievable in the geometrical range, which corresponds to \aposteriori distribution, $p(\, \boldsymbol{r}_{(1,2,\ldots,n)} \,|\, \Delta \boldsymbol{N},\, \boldsymbol{y}_{(1,2,\ldots,n)} )$, (see also Appendix~II). It is calculated under the condition~(a) in Table~II. This investigation is important because the actual precise positioning should be evaluated from the perspectives on its attainable geometrical precision in combination with the performance of integer ambiguity resolution.

\putfloat[t]{temporal variations} %

The error variances attained from the L1, L1+L2, and L1+L2+L5 measurements retain the original precision in carrier-phase measurements and achieve the high precision in the geometrical range except for the L1 measurements under the condition of large $\sigma_{\varDelta I}$. Those attained from the $\text{L}_\text{W}$ and $\text{L}_\text{W}$+$\text{L}_\text{EW}$ measurements, by contrast, lose the original precision in carrier-phase measurements and have a larger range error than that from the L1 measurements. This reveals the trade-off in the use of these artificial measurements ($\text{L}_\text{W}$ and $\text{L}_\text{W}$+$\text{L}_\text{EW}$) between the precision in positioning and the performance of integer ambiguity resolution under the condition of large $\sigma_{\varDelta I}$.
\subsection{Temporal variations} %

We investigated the dependence of the optimal success rate on both time and sampling duration, the latter of which is the initialization time consumed in the integer ambiguity resolution.

Figs.~\ref{temporal variations L1+L2+L5} and \ref{temporal variations L1+L2} show the temporal variations of the success rate, $\alpha_\mathrm{MAP}$, under the conditions of the use of L1+L2+L5 and L1+L2 measurements, respectively, and under the distinct conditions of the sampling duration, which is set to 1, 10, 25, or 50 seconds and depicted as horizontal bars of corresponding length. The satellites' constellation and its movement is derived by using a GPS broadcast ephemeris in this calculation. The error variance of single-differentiated ionospheric delay, $\sigma_{\varDelta I}$, is set to 5\,mm and other parameters are set to the same as those under the distinct conditions~(a) in Table~II and the common ones in Table~III.

The two distinct conditions of the time-constant receiver's coordinates (static ambiguity resolution) and the non-time-correlative ones (kinematic one) are additionally examined and depicted in each figure except for the sampling durations of one second (note that the measurement rate is 1\,epoch/sec in this calculation). Each plot of $\alpha_\mathrm{MAP}$ is essentially calculated by using the Monte Carlo integration.

The results indicate that the constellation of the satellites affects $\alpha_\mathrm{MAP}$ whose temporal variations are estimated to have almost the same amplitude of about 0.5 in $\log_{10} \alpha_\mathrm{MAP} / ( 1 - \alpha_\mathrm{MAP} )$ for each duration, appearing in the whole calculated time (300 seconds). The rate grows with a longer sampling duration, and, furthermore, the growth is accelerated by the constellation's movement within the duration. The acceleration effect is, however, not temporally homogeneous (compare the plot of the 50-second duration with the one of 1-second, in regard to the degree of decrease after 200 seconds in both the figures).

The effect of the satellites' constellation largely affects $\alpha_\mathrm{MAP}$ especially under the conditions of the use of L1+L2 measurements and the short (1-second) sampling duration shown in Fig.~\ref{temporal variations L1+L2}, which contains the short-term degradation in the success rate (compare the plot at about 120 seconds with the one under the use of L1+L2+L5 measurements in Fig.~\ref{temporal variations L1+L2+L5}).

The static ambiguity resolution has slightly higher success rates than those of the kinematic one, as shown in both the figures. This is caused by the constellation's slight movement during the ambiguity resolution. The effect is not temporally homogeneous either, as also shown in the figures.

\section{Conclusion} %

This study developed the almost rigorous calculations of the maximum-marginal-\textit{a-posteriori} success rates of statistical decision for multivariate Gaussian error distribution on integer lattices, and applied them to revealing the various possible realizations of the reliable and short-period integer ambiguity resolution in carrier-phase relative positioning by GPS/GNSS. 

The extremely high reliability was shown under the conditions of accurate multiple-frequency carrier-phase measurements and a large number of visible satellites. Its heavy degradation was revealed under the conditions of the even slight amount of differentiated ionospheric delays originating from the nonvanishing baseline length between the rover and reference receivers. The succeeding initialization of the integer ambiguities was shown to overcome such a disadvantageous condition remarkably due to the temporal and spatial constancy of differentiated ionospheric delays. The slight difference was also shown in the performance between the static and kinematic short-period ambiguity resolutions.

This study makes it possible to investigate the achievable performance of the short-term integer ambiguity resolution in the field of GPS/GNSS rigorously.

\appendices

\section{}  %

A uniform, or translationally invariant (location-invariant), infinite-interval (\apriori{}) distribution should be carefully defined to be the limit of a finite-interval distribution:
\begin{equation}
p(\theta) = \lim_{L \rightarrow \infty} p_L (\theta)
,
\nonumber
\end{equation}
where
\begin{equation}
p_L (\theta) = \frac{1}{2 L} \: ; \mspace{14mu} -L \leq \theta \leq L
,
\nonumber
\end{equation}
in the case of a continuous distribution, or
\begin{equation}
p_L (\theta)= \frac{1}{2 L + 1} \: ; \mspace{14mu} \theta = -L,\,-L+1,\,\ldots,\,L
,
\nonumber
\end{equation}
in the case of an integer-valued distribution.
The limit should be executed at the final process of the calculation of the marginal \aposteriori{} distribution in order to avoid the indeterminate division which may be caused by the improper normalization coefficient.
\section{}  

Given $\Delta \boldsymbol{N}$ and $\boldsymbol{y}_{(1,2,\ldots,n)}$, the marginal \aposteriori{} distribution of $\boldsymbol{I}_{(1,2,\ldots,n)}$, or $p(\, \boldsymbol{I}_{(1,2,\ldots,n)} \,|\, \Delta \boldsymbol{N},\, \boldsymbol{y}_{(1,2,\ldots,n)} )$, is derived by integrations and the use of the Bayes theorem as follows:
\begin{align}
& p(\, \boldsymbol{I}_{(1,2,\ldots,n)} \,|\, \Delta \boldsymbol{N},\, \boldsymbol{y}_{(1,2,\ldots,n)} )
\nonumber\\
& = \int p(\, \boldsymbol{I}_{(1,2,\ldots,n)},\, \Delta \boldsymbol{\tau}_{(1,2,\ldots,n)} \,|\, \Delta \boldsymbol{N},\, \boldsymbol{y}_{(1,2,\ldots,n)} )
\nonumber\\
& \hspace{16.5em} \times d \Delta \boldsymbol{\tau}_{(1,2,\ldots,n)} ,
\nonumber\\
& p(\, \boldsymbol{I}_{(1,2,\ldots,n)},\, \Delta \boldsymbol{\tau}_{(1,2,\ldots,n)} \,|\, \Delta \boldsymbol{N},\, \boldsymbol{y}_{(1,2,\ldots,n)} )
\nonumber\\
& = \int p(\, \boldsymbol{r}_{(1,2,\ldots,n)},\, \boldsymbol{I}_{(1,2,\ldots,n)},\, \Delta \boldsymbol{\tau}_{(1,2,\ldots,n)} \,|\, \Delta \boldsymbol{N},\, 
\nonumber\\
& \hspace{14em} \boldsymbol{y}_{(1,2,\ldots,n)} ) \ d \boldsymbol{r}_{(1,2,\ldots,n)} ,
\nonumber\\
& p(\, \boldsymbol{r}_{(1,2,\ldots,n)} \,|\, \Delta \boldsymbol{N},\, \boldsymbol{y}_{(1,2,\ldots,n)} )
\nonumber\\
& = \int p(\, \boldsymbol{r}_{(1,2,\ldots,n)},\, \boldsymbol{I}_{(1,2,\ldots,n)},\, \Delta \boldsymbol{\tau}_{(1,2,\ldots,n)} \,|\, \Delta \boldsymbol{N},\, 
\nonumber\\
& \hspace{8.5em} \boldsymbol{y}_{(1,2,\ldots,n)} ) \ d \boldsymbol{I}_{(1,2,\ldots,n)} d \Delta \boldsymbol{\tau}_{(1,2,\ldots,n)},
\nonumber\\
& p(\, \boldsymbol{r}_{(1,2,\ldots,n)},\, \boldsymbol{I}_{(1,2,\ldots,n)},\, \Delta \boldsymbol{\tau}_{(1,2,\ldots,n)} \,|\, \Delta \boldsymbol{N},\, \boldsymbol{y}_{(1,2,\ldots,n)} )
\nonumber\\
& = \iint \sum_{\boldsymbol{N}^0} p(\, \boldsymbol{N}^0,\, \boldsymbol{r}_{(1,2,\ldots,n)},\, \boldsymbol{I}_{(1,2,\ldots,n)},\, \Delta \boldsymbol{\tau}_{(1,2,\ldots,n)},
\nonumber\\
& \hspace{6em} \tau_{(1,2,\ldots,n)}^0,\, \psi_{(1,2,\ldots,n)} \,|\, \Delta \boldsymbol{N},\, \boldsymbol{y}_{(1,2,\ldots,n)} )
\nonumber\\
& \hspace{13em}
\times d \tau_{(1,2,\ldots,n)}^0 d \psi_{(1,2,\ldots,n)} ,
\nonumber\\
& p(\, \boldsymbol{N}^0,\, \boldsymbol{r}_{(1,2,\ldots,n)},\, \boldsymbol{I}_{(1,2,\ldots,n)},\, \boldsymbol{\tau}_{(1,2,\ldots,n)},\, \psi_{(1,2,\ldots,n)} \,|
\nonumber\\
& \hspace{16em} \Delta \boldsymbol{N},\, \boldsymbol{y}_{(1,2,\ldots,n)} )
\nonumber\\
& = p(\, \boldsymbol{N}^0,\, \boldsymbol{r}_{(1,2,\ldots,n)},\, \boldsymbol{I}_{(1,2,\ldots,n)},\, \boldsymbol{\tau}_{(1,2,\ldots,n)},\, \psi_{(1,2,\ldots,n)},
\nonumber\\
& \hspace{7.4em} \boldsymbol{y}_{(1,2,\ldots,n)} \,|\, \Delta \boldsymbol{N} ) \ / \ p(\, \boldsymbol{y}_{(1,2,\ldots,n)} \,|\, \Delta \boldsymbol{N} ) ,
\nonumber\\
& p(\, \boldsymbol{y}_{(1,2,\ldots,n)} \,|\, \Delta \boldsymbol{N} )
\nonumber\\
& = \iiiint \sum_{\boldsymbol{N}^0} p(\, \boldsymbol{N}^0,\, \boldsymbol{r}_{(1,2,\ldots,n)},\, \boldsymbol{I}_{(1,2,\ldots,n)},\, \boldsymbol{\tau}_{(1,2,\ldots,n)},\, 
\nonumber\\
& \hspace{12em} \psi_{(1,2,\ldots,n)},\, \boldsymbol{y}_{(1,2,\ldots,n)} \,|\, \Delta \boldsymbol{N} )
\nonumber\\
& \hspace{4.2em} \times d \boldsymbol{r}_{(1,2,\ldots,n)} d \boldsymbol{I}_{(1,2,\ldots,n)} d \boldsymbol{\tau}_{(1,2,\ldots,n)} d \psi_{(1,2,\ldots,n)} ,
\nonumber\\
& p(\, \boldsymbol{N}^0,\, \boldsymbol{r}_{(1,2,\ldots,n)},\, \boldsymbol{I}_{(1,2,\ldots,n)},\, \boldsymbol{\tau}_{(1,2,\ldots,n)},\, 
\nonumber\\
& \hspace{12em} \psi_{(1,2,\ldots,n)},\, \boldsymbol{y}_{(1,2,\ldots,n)} \,|\, \Delta \boldsymbol{N} )
\nonumber\\
& = p(\, \boldsymbol{y}_{(1,2,\ldots,n)}\,|\, \Delta \boldsymbol{N},\, \boldsymbol{N}^0,\, \boldsymbol{r}_{(1,2,\ldots,n)},\, \boldsymbol{I}_{(1,2,\ldots,n)},
\nonumber\\
& \hspace{14.5em} \boldsymbol{\tau}_{(1,2,\ldots,n)},\, \psi_{(1,2,\ldots,n)})
\nonumber\\
& \hspace{1em} \times p(\, \boldsymbol{N}^0,\, \boldsymbol{r}_{(1,2,\ldots,n)},\, \boldsymbol{I}_{(1,2,\ldots,n)},\, \boldsymbol{\tau}_{(1,2,\ldots,n)},\, \psi_{(1,2,\ldots,n)}) ,
\nonumber\\
& \boldsymbol{\tau}_{(1,2,\ldots,n)} = \Delta \boldsymbol{\tau}_{(1,2,\ldots,n)} + (1,\,1,\,\ldots,\,1)^T \times \tau_{(1,2,\ldots,n)}^0 ,
\nonumber
\end{align}
where \apriori{} distribution, 
\begin{equation}
p(\, \boldsymbol{N}^0,\, \boldsymbol{r}_{(1,2,\ldots,n)},\, \boldsymbol{I}_{(1,2,\ldots,n)},\, \boldsymbol{\tau}_{(1,2,\ldots,n)},\, \psi_{(1,2,\ldots,n)}) , 
\nonumber
\end{equation}
is used.

\bibliographystyle{IEEEtran}
\bibliography{IEEEabrv,myabbrev,mybib}

\begin{savefloat}{voronoi polytopes}
\begin{figure}[t]
\begin{center}
\subfigure[$\Delta \boldsymbol{N}$-separable case in the space of $\Delta \boldsymbol{\nu}$]{
\includegraphics[width=45mm]{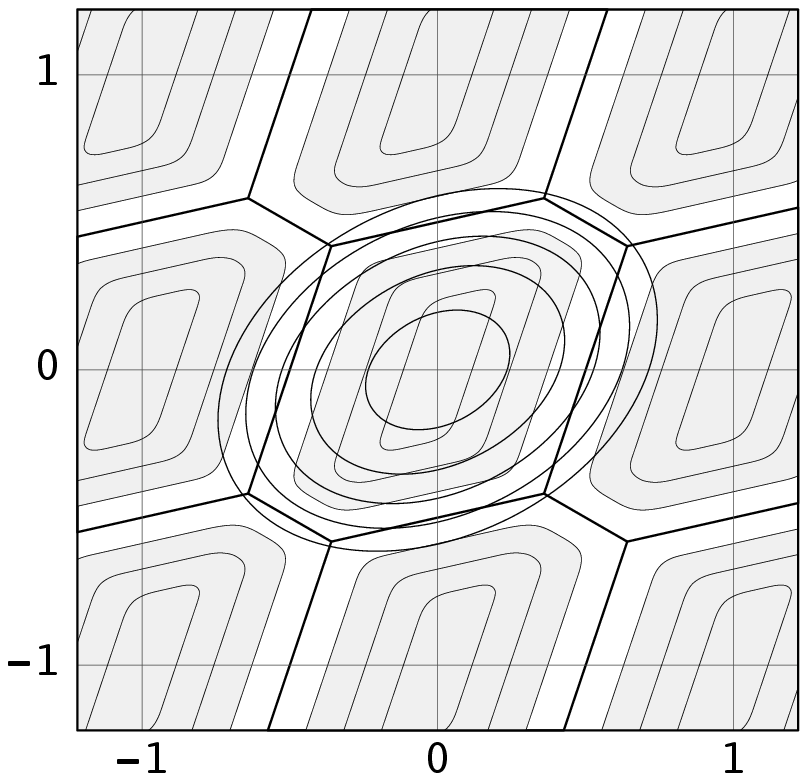}
\hspace{2.0mm}
\label{voronoi polytope in separable case}
}
\\ %
\subfigure[$\Delta \boldsymbol{N}$-nonseparable case in the space of $\boldsymbol{\nu}$]{
\includegraphics[width=45mm]{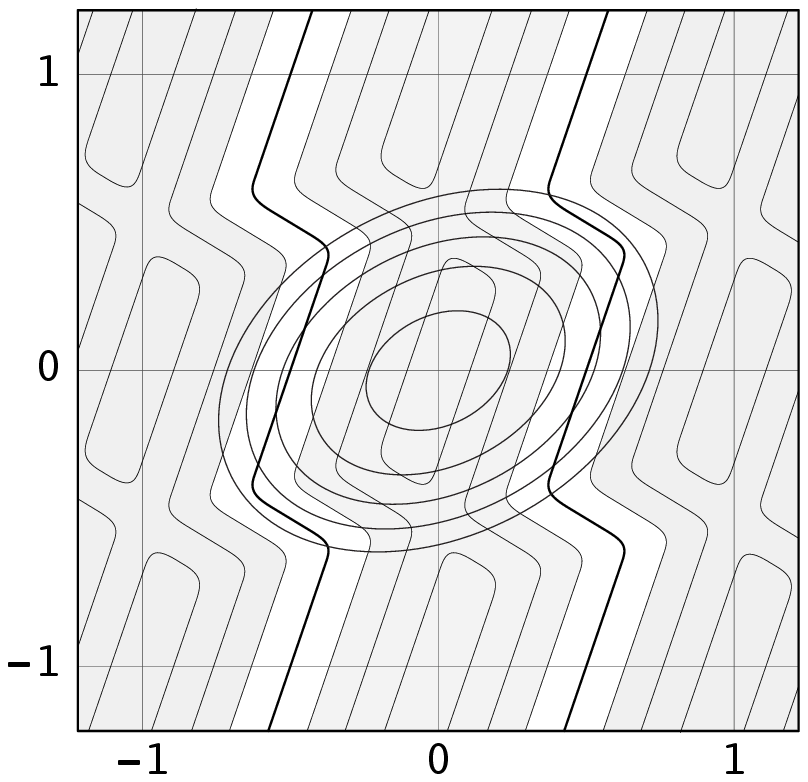}
\hspace{2.0mm}
\label{voronoi polytope in nonseparable case}
}
\caption{Two-dimensional error distributions and the acceptance regions of $D_\mathrm{MAP} (\Delta \boldsymbol{N})$ and $D (\Delta \boldsymbol{N}; h)$ for success/error rates.}
\label{voronoi polytopes}
\end{center}
\end{figure}
\end{savefloat}%
\begin{savefloat}{errorintegral}
\begin{figure}\centering
\centerline {
\hspace{4.0mm}
\includegraphics[width=65mm]{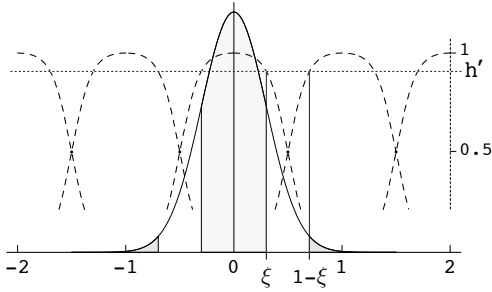}
}
\caption{One-dimensional error distribution and the acceptance regions for success/error rates.}
\label{errorintegral}
\end{figure}
\end{savefloat}%
\begin{savefloat}{conditional rates in separable case}
\begin{figure}[t]
\begin{center}
\subfigure[success rate]{
\includegraphics[width=78mm]{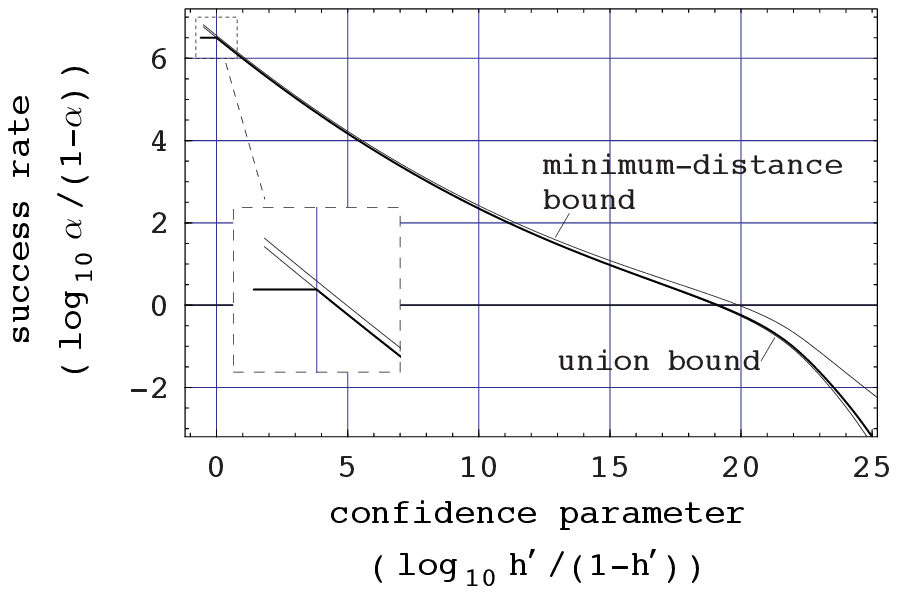}
\hspace{2.3em}
\label{success rate in separable case}
}
\\ %
\subfigure[error rate]{
\includegraphics[width=78mm]{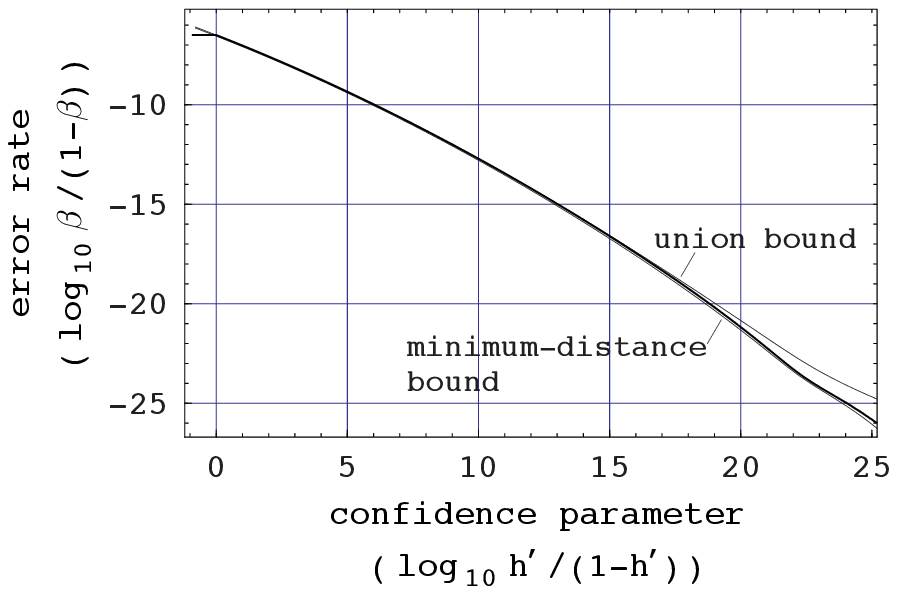}
\hspace{2.0em}
\label{error rate in separable case}
}
\caption{The rates calculated by using Monte Carlo integration and approximated bounds in the $\Delta \boldsymbol{N}$-separable case.}
\label{conditional rates in separable case}
\end{center}
\end{figure}
\end{savefloat}%
\begin{savefloat}{conditional rates in nonseparable case}
\begin{figure}[t]
\begin{center}
\subfigure[success rate]{
\includegraphics[width=78mm]{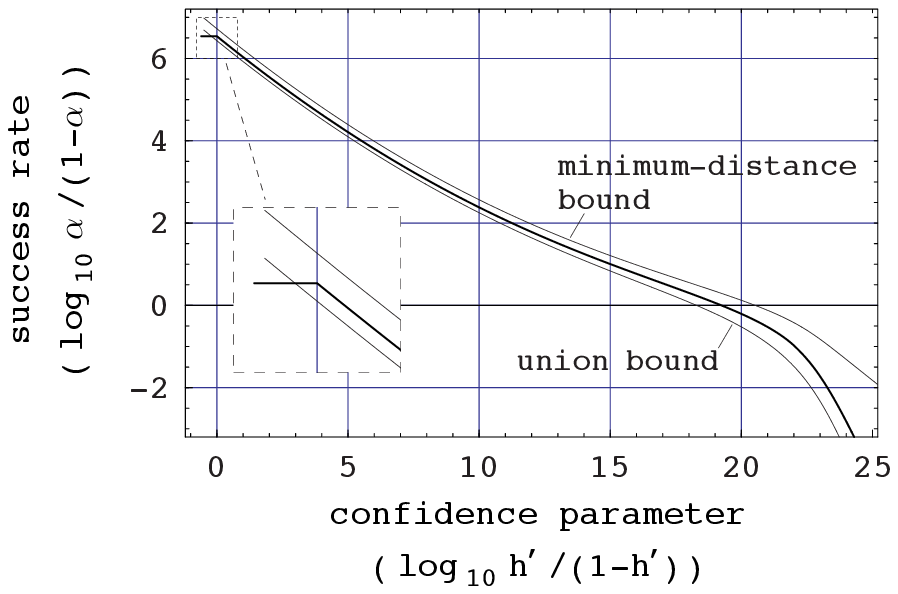}
\hspace{2.3em}
\label{success rate in nonseparable case}
}
\\ %
\subfigure[error rate]{
\includegraphics[width=78mm]{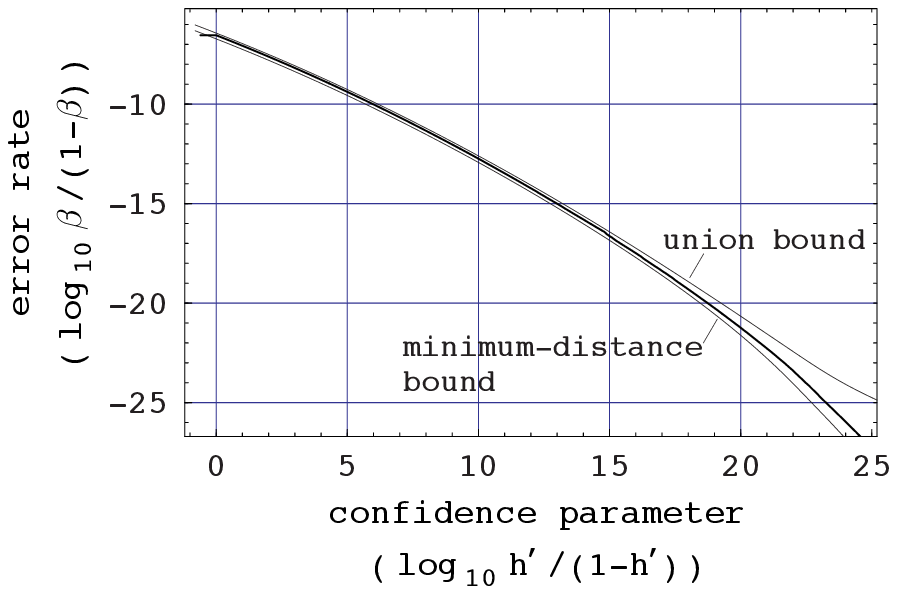}
\hspace{2.0em}
\label{error rate in nonseparable case}
}
\caption{The rates calculated by using Monte Carlo integration and approximated bounds in the $\Delta \boldsymbol{N}$-nonseparable case.}
\label{conditional rates in nonseparable case}
\end{center}
\end{figure}
\end{savefloat}%
\begin{savefloat}{dependence of success rates}
\begin{figure}[t]
\begin{center}
\subfigure[condition~(a) in Table~II]{
\includegraphics[width=80mm]{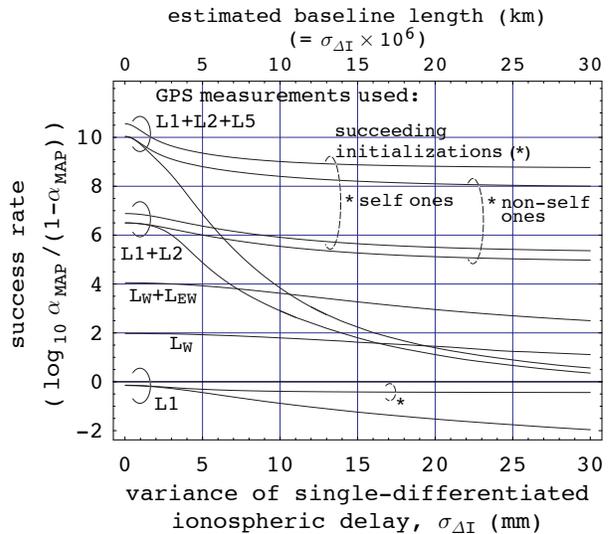}
\hspace{1.8em}
\label{dependence of success rates a}
}
\\ %
\subfigure[condition~(b) in Table~II]{
\includegraphics[width=80mm]{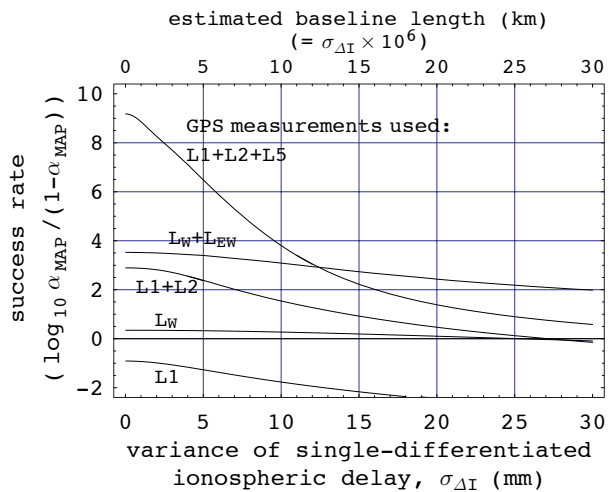}
\hspace{1.5em}
\label{dependence of success rates b}
}
\caption{The dependence of the success rates of integer ambiguity resolution on the uncertainty of ionospheric delay.}
\label{dependence of success rates}
\end{center}
\end{figure}
\end{savefloat}%
\begin{savefloat}{range-error dependence}
\begin{figure}\centering
\centerline {
\includegraphics[width=80mm]{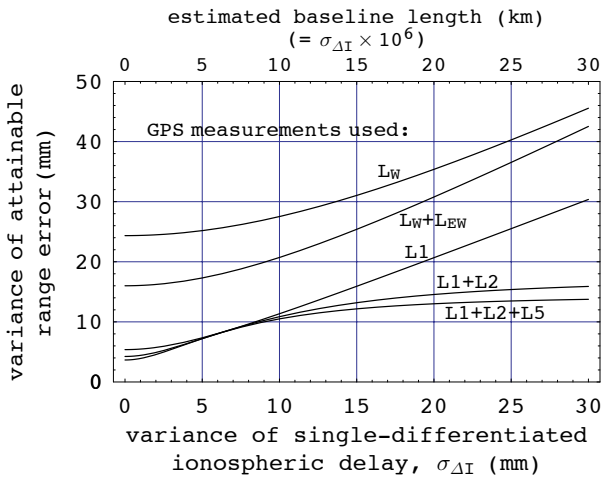}
\hspace{1.5em}
}
\caption{Range-error dependence on the uncertainty of ionospheric delay.}
\label{dependence of range error}
\end{figure}
\end{savefloat}%
\begin{savefloat}{temporal variations}
\begin{figure}[t]
\begin{center}
\subfigure[L1+L2+L5 measurements]{
\includegraphics[width=80mm]{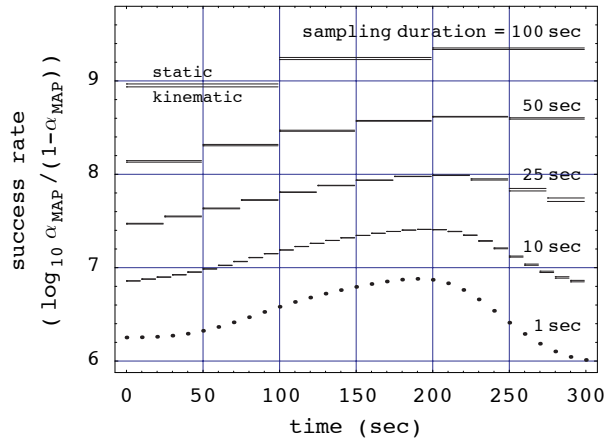}
\hspace{3.4mm}
\label{temporal variations L1+L2+L5}
}
\\ %
\subfigure[L1+L2 measurements]{
\includegraphics[width=80mm]{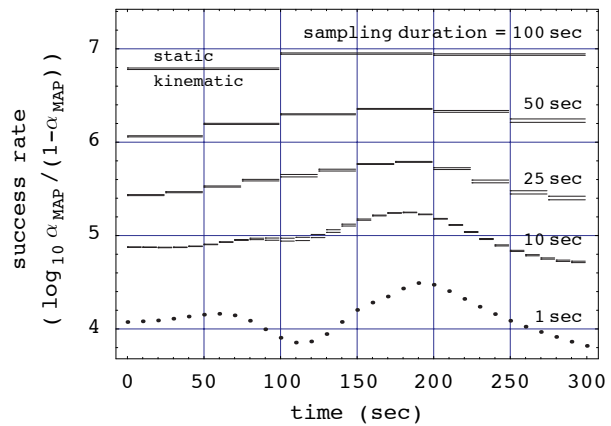}
\hspace{3.2mm}
\label{temporal variations L1+L2}
}
\caption{Temporal variations of the success rates of integer ambiguity resolution under the condition of several sampling durations and the use of two distinct measurement sets.}
\label{temporal variations}
\end{center}
\end{figure}
\end{savefloat}
\begin{savefloat}{distinct_sets_measurements}
\begin{table}[t] %
\caption{The distinct sets of GPS measurements used in the calculations.}
\label{distinct_sets_measurements}
\begin{center}
\begin{tabular}{ll}
\Hline
symbol & the set of measurements \\
\hline
L1 & $\phi_{\text{L1}}$ and $\rho_{\text{L1}}$ \rule{0pt}{2.4ex} \\

L1+L2 & $\phi_{\text{L1}}$, $\phi_{\text{L2}}$, $\rho_{\text{L1}}$, and $\rho_{\text{L2}}$ \\

L1+L2+L5 & $\phi_{\text{L1}}$, $\phi_{\text{L2}}$, $\phi_{\text{L5}}$, $\rho_{\text{L1}}$, $\rho_{\text{L2}}$, and $\rho_{\text{L5}}$ \\

$\text{L}_\text{W}$ & $\phi_{\text{L}_\text{W}}$, $\rho_{\text{L1}}$, and $\rho_{\text{L2}}$ \\

$\text{L}_\text{W}$+$\text{L}_\text{EW}$ & $\phi_{\text{L}_\text{W}}$, $\phi_{\text{L}_\text{EW}}$, $\rho_{\text{L1}}$, $\rho_{\text{L2}}$, and $\rho_{\text{L5}}$ \vspace{0.1em} \\
\Hline
\end{tabular}
\end{center}
\end{table}
\end{savefloat}%
\begin{savefloat}{parameters_ Individual}
\begin{table}[t] %
\caption{Individual conditions of calculations.}
\label{parameters_ Individual}
\begin{center}
\begin{tabular}{llll}
\Hline
con- & \multicolumn{3}{l}{variance of error in single-differentiated} \rule{0pt}{2.0ex} \\
dition & \multicolumn{3}{l}{code-pseudorange measurements} \vspace{0.3em} \\
 & $\bullet$ time-varying & $\bullet$ time-constant \\
 & \phantom{$\bullet$} component & \phantom{$\bullet$} component \\
\hline
(a) & \phantom{$\bullet$} 0.5\,m & \phantom{$\bullet$} 0.5\,m \rule{0pt}{2.3ex} \\
(b) & \phantom{$\bullet$} 2.0\,m & \phantom{$\bullet$} 2.0\,m \\

\Hline
\end{tabular}
\end{center}
\end{table}
\end{savefloat}
\begin{savefloat}{parameters_common}
\begin{table}[t] %
\caption{Common parameters of the calculations.}
\label{parameters_common}
\begin{center}
\begin{tabular}{ll}
\Hline
number of GPS satellites & 7 \rule{0pt}{2.2ex} \\
variance of error in single-differenti- \\
ated carrier-phase measurements\\
\quad$\bullet$ time-varying component & 0.02\,cycle \\
\quad$\bullet$ time-constant component & 0.02\,cycle \vspace{0.3em} \\
AR(1) coefficient in time-varying \\
component \\
\hspace{0.7em}$\bullet$ carrier-phase measurements & 0.95 \\
\hspace{0.7em}$\bullet$ code-pseudorange measurements & 0.5 \vspace{0.3em} \\
measurement rate & 1\,epoch/sec \vspace{0.3em} \\
measurement time & 10\,sec \vspace{0.3em} \\
pre-measurement time & 10\,sec\\
in the succeeding initialization case \vspace{0.3em} \\
\apriori{} prediction of receiver's & non-informative and\\%
 coordinates and clock offset& non-time-correlative \vspace{0.3em} \\
\apriori{} prediction of receiver's & non-informative and \\
interfrequency biases & time-constant \vspace{0.3em} \\
\apriori{} prediction of single- & predicted with uncer-\\
differentiated ionospheric delay ($\sigma_{\Delta I}$) & tainty and time-con-\\
 & stant \\
\apriori{} prediction of single- & 0 (deterministic) \\
differentiated tropospheric delay & \\
\Hline
\end{tabular}
\end{center}
\end{table}
\end{savefloat}
\end{document}